\documentclass[usenatbib,preprint]{mn2ePASJ}

\usepackage[switch,mathlines]{lineno}
\newcommand*\patchAmsMathEnvironmentForLineno[1]{%
  \expandafter\let\csname old#1\expandafter\endcsname\csname #1\endcsname
  \expandafter\let\csname oldend#1\expandafter\endcsname\csname end#1\endcsname
  \renewenvironment{#1}%
     {\linenomath\csname old#1\endcsname}%
     {\csname oldend#1\endcsname\endlinenomath}}%
\newcommand*\patchBothAmsMathEnvironmentsForLineno[1]{%
  \patchAmsMathEnvironmentForLineno{#1}%
  \patchAmsMathEnvironmentForLineno{#1*}}%
\AtBeginDocument{%
\patchBothAmsMathEnvironmentsForLineno{equation}%
\patchBothAmsMathEnvironmentsForLineno{align}%
\patchBothAmsMathEnvironmentsForLineno{flalign}%
\patchBothAmsMathEnvironmentsForLineno{alignat}%
\patchBothAmsMathEnvironmentsForLineno{gather}%
\patchBothAmsMathEnvironmentsForLineno{multline}%
}

\usepackage{graphicx}

\usepackage {natbib,aas_macros}
\usepackage {mediabb}
\usepackage {bm}
\usepackage {amsmath,amssymb}
\usepackage {setspace}

\newcommand{\msun}{\thinspace M_\odot}
\newcommand{\msunyr}{\thinspace M_\odot\thinspace {\rm yr^{-1}}} 
\newcommand{\gcm}{~{\rm g~cm}^{-3} }
\newcommand{\gcmcm}{~{\rm g~cm}^{-2} }

\newcommand{\magB}{\mathbf{B}}

\newcommand{\cms}{~{\rm cm} ~{\rm s}^{-1} } 
\newcommand{\cmcms}{~{\rm cm}^2 ~{\rm s}^{-1} }

\newcommand{\mum}{{\rm \mu} {\rm m} }

\newcommand{\gauau}{{\rm G~ AU^2}}

\newcommand{\mm}{{\rm mm}}

\newcommand{\jEnv}{\left(\frac{j_{\rm env}}{10^{-3} {\rm ~km~ pc~ s^{-1}}}\right)}
\newcommand{\massDisk}{\left(\frac{M_{\rm star}}{0.3 \msun}\right)}
\newcommand{\tempDisk}{\left(\frac{T_0}{150 ~{\rm K}}\right)}
\newcommand{\mdotDisk}{\left(\frac{\dot{M}_{\rm disk}}{10^{-6} ~\msun {\rm yr}^{-1}}\right)}
\newcommand{\zetaCRDisk}{\left(\frac{\zeta}{10^{-18} ~{\rm s^{-1}}}\right)}
\newcommand{\radDiskSize}{\left(\frac{r_{\rm disk}}{100 ~{\rm AU}}\right)}
\newcommand{\radDisk}{\left(\frac{r}{10 ~{\rm AU}}\right)}

\title{Co-evolution of dust grains and protoplanetary disks II: structure and evolution of protoplanetary disks; an analytical approach}

\author[Tsukamoto et al]{
Yusuke Tsukamoto$^{1}$ \\
$^1$Graduate Schools of Science and Engineering, Kagoshima University, Kagoshima, Japan  \\
}

\begin{document}
\maketitle

\begin{abstract}
  In our previous study \citep{2023PASJ...75..835T}, we investigated formation and early evolution of protoplanetary disks with 3D non-ideal magnetohydrodynamics simulations
  considering dust growth, and found that the modified equations of the conventional steady accretion disk model which consider the magnetic braking, { dust growth} and ambipolar diffusion
  reproduce the disk structure (such as density and vertical magnetic field) obtained from simulations very well.
  The disk structure predicted from the disk model is described by observationally or experimentally constrainable parameters
  such as the  mass of central star, mass accretion rate, recombination rate, temperature, and ionization rate.
  In this paper, as a sequel of the our previous study, 
  we analytically investigate the structure and evolution of protoplanetary disks corresponding to Class 0/I young stellar objects
  using the modified steady accretion disk model combining an analytical model of envelope accretion.
  We estimate that the disk radius is several AU at disk formation epoch and increases to several 100 AU at the end of the accretion phase.
  The disk mass is estimated to be $0.01 \msun \lesssim M_{\rm disk} \lesssim 0.1 \msun$ for a disk with radius of several 10 AU and
  mass accretion rate of $\dot{M}_{\rm disk} \sim 10^{-6} \msunyr$. 
  These estimates seems to be consistent with recent observations.
  We also found that, with typical disk ionization rates ($\zeta \gtrsim 10^{-19}  {\rm ~s^{-1}}$) and moderate mass accretion rate ($\dot{M}_{\rm disk}\gtrsim10^{-8} \msunyr$),
  magneto-rotational instability is suppressed in the disk because of low plasma $\beta$ and efficient ambipolar diffusion.
  We argue that the radial profile of specific angular momentum (or rotational velocity) at the disk outer edge should be continuously connected to that of the envelope if the disk evolves by magnetic braking, and should be discontinuous if the disk evolves by internal angular momentum transport process such as gravitational instability or magneto-rotational instability. 
  Future detailed observations of the specific angular momentum profile around the disk outer edge
  are important for understanding the angular momentum transport mechanism of protoplanetary disks. 
\end{abstract}


\begin{keywords}
star formation -- circum-stellar disk -- methods: magnetohydrodynamics -- protoplanetary disk
\end{keywords}

\section{Introduction}
\label{intro}
Predicting the structure and  evolution of protoplanetary disks such as surface density and magnetic field profile
with observationally or experimentally constrainable parameters is a major challenge in theoretical astrophysics.
Toward this end, many studies have been carried out using multidimensional magnetohydrodynamics (MHD)
simulations that take into account the non-ideal MHD effects \citep{
  2011ApJ...738..180L, 2011PASJ...63..555M,2015ApJ...801..117T,2015MNRAS.452..278T,2015ApJ...810L..26T,
  2016MNRAS.457.1037W,2016A&A...587A..32M,2017PASJ...69...95T,
  2018MNRAS.473.4868Z,2019MNRAS.486.2587W,2020ApJ...896..158T,2020MNRAS.492.3375Z,2021MNRAS.508.2142X}.
The consensus from these studies is that non-ideal MHD effects, especially ambipolar diffusion, play a crucial role in disk evolution.
Because the strength of non-ideal MHD effects, magnetic resistivity strongly depends on the microscopic physics
(such as the cosmic-ray ionization, the gas-phase recombination, and the adsorption rate of charged particles on the dust grains),
various studies have also been carried out to investigate the impact of the microscopic physics on the disk evolution
\citep{2016MNRAS.460.2050Z,2018MNRAS.tmp..378W,2020ApJ...896..158T, 2020A&A...639A..86K,2021A&A...649A..50M,2023A&A...670A..61M,2023MNRAS.tmp..691K}.

While it is wonderful that such multidimensional, microphysics-aware simulations are now possible and thereby reveal the formation and evolution of protoplanetary disks,
the high computational costs of such simulations make it difficult to study the long-term evolution of protoplanetary disks over $\sim 10^6$ yr after protostar formation, through the Class 0 and Class I evolutionary stages, to the Class II stage.
Furthermore, their microscopic and macroscopic complexities make it seem extremely difficult to achieve the ultimate goal described above, i.e.,
to concisely describe the disk evolution and disk structure in terms of a few parameters.

Recently, \citet{2023PASJ...75..835T} (here after TMI23) investigated the formation and evolution of protoplanetary disks
using dust-gas two-fluid non-ideal MHD simulations with dust growth.
The simulations also take into account the changes of magnetic resistivity associated with dust growth.
This study proposed the concept of a "co-evolution of dust grains and protoplanetary disks",
a process in which the global evolution of the disk is significantly affected by dust growth
through the changes in the adsorption efficiency of charged particles onto the dust grains.

TMI23 also found that when the dust grains becomes sufficiently large (broadly speaking, the maximum dust size becomes $100 ~\mum$ to $1 ~\mm$; see \citet{2022ApJ...934...88T} for more detailed calculations) and the adsorption of ions and electrons on the dust grains becomes negligible,
the radial profiles of the disk such as density, magnetic field, radial velocity, and ambipolar resistivity are well described by the power laws.
TMI23 also showed that a simple extension of the steady accretion disk model reproduces the disk structure surprisingly well not only the power law exponent but also the absolute value.

The disk structure of TMI23 is described in terms of observationally constrainable (or experimentally measurable, such as the gas phase recombination rate) parameters,
and is conducive to the ultimate goal described above.
Furthermore, because it is written analytically, it can be an important tool for understanding long term evolution of the disks until the end of Class I.
Therefore, it is important to clarify the predictions obtained from the model
and give suggestions for future observations.

However, the disk structure derived in TMI23 is not suitable for the purpose of comparison with observations
because the model employed (simulation-based) assumptions such as barotropic equation of state to allow the detailed comparisons with the simulation results.
Therefore, in this paper, we derive a disk structure suitable for comparison to observations
by making simple assumptions on the temperature structure and gas-phase recombination rate for the disk model proposed in TMI23. Then we discuss its properties and,
by combining an analytical model of envelope accretion and the disk model,
we make observationally testable predictions. 
We think that some of the nontrivial predictions for disk mass, radius,
and specific angular momentum profile presented in this paper can be verified by future observations.

\section{Governing equations and the disk structure}
\subsection{Governing equations of disk model}
In TMI23, we suggested that the disk structures obtained by non-ideal MHD simulations
are well described by a steady accretion disk model of,
\begin{eqnarray}
  \label{eq1}
  -2 \pi r v_r \Sigma =\dot{M}_{\rm disk},\\
  \label{eq2}
  v_\phi =\sqrt{\frac{G M_{\rm star}}{r}}\equiv r \Omega,\\
  \label{eq3}
  H=\frac{c_s}{\Omega},\\
  \label{eq4}
   \Sigma (r \Omega) v_r =- r \frac{B_z B_{\phi, s}}{\pi},\\
  \label{eq5}
  B_z v_r=-\frac{\eta_A}{r} B_z,\\
  \label{eq6}
  \eta_A \frac{B_{\phi, s}}{H}  =\left( \frac{H}{r}\right)^2 B_z v_\phi.
\end{eqnarray}
where $r$ is the radius from the central protostar,
$\Sigma$ is the surface density of the disk,
$v_r$ is the radial velocity of the gas,
$\dot{M}_{\rm disk}$ is the mass accretion rate in the disk which is assumed to be constant,
$v_\phi$ is the azimuthal velocity of the gas,
$\Omega$ is the angular velocity.
$M_{\rm star}$ is the protostar mass,
$c_s$ is the sound speed,
$H$ is the scale height,
$B_z$ is the vertical magnetic field,
$B_{\phi, s}$ is the azimuthal magnetic field at $z=H$,
$\eta_A$ is the ambipolar resistivity.
{ Among these quantities, $\dot{M}_{\rm disk}$ and $M_{\rm star}$ are constants and do not depend on the radius
  and other quantities are functions of the radius.
}
  $G$ is the gravitational constant.

The equation (\ref{eq1}) to equation (\ref{eq3}) are the same as the standard viscous accretion disk model and our extension is expressed in (\ref{eq4}) to (\ref{eq6}).
The equation (\ref{eq4})  shows that the magnetic braking balances the radial angular momentum advection.
The equation (\ref{eq5})  shows that the radial magnetic flux advection balances the radial magnetic field drift by the ambipolar diffusion.
The equation (\ref{eq6})  shows that the azimuthal magnetic field generation by the vertical shear motion
balances the azimuthal magnetic field drift by the ambipolar diffusion.

Strictly speaking, TMI23 showed that the solutions of equations (A29) to (A33) in TMI23
which are derived from perturbation theory reproduce the simulation results well.
However, we have confirmed that the disk structures predicted by the above equations and those predicted by the equations used in TMI23 are essentially the same in absolute values as well as in power exponents
(for example, the differences in values of $\rho$ and $B_z$ are about 20\%, and the all power exponents of the physical quantities are completely the same).
So, we use the equations above in this paper because they are simpler and more intuitive.

To obtain the solution of the equations (\ref{eq1}) to (\ref{eq6}), we need a model of $\eta_A$ and the temperature profile.
We showed in \citet{2022ApJ...934...88T} that, as dust growth proceeds, $\eta_A$ tends to decrease and converges to the analytic power law of
\begin{align}
  \label{model_etaA}
  \eta_A&=\frac{\magB^2}{4 \pi C \gamma \rho^{3/2}} \sim \frac{B_z^2}{4 \pi C \gamma \rho_{\rm mid}^{3/2}},
\end{align}
even in the disk, which is determined by the balance between gas phase ionization and gas phase recombination \citep{1983ApJ...273..202S}.
{ Here, $\rho_{\rm mid}$ is the midplane density of the disk.}
{ To derive the right-hand side expression, we assume that the vertical magnetic field is dominant and the contribution of the radial and azimuthal magnetic field on the field strength are negligible.
  We also assume that the density variation in the vertical direction is negligible.}

Note that, using this formula for $\eta_A$, we model that the dust grains have sufficiently grown in the disk.
{
  Broadly speaking, this formula is justified if the maximum dust size becomes $a_{\rm max}\gtrsim 100 \mum$ in the disk 
  \citep[see,][for more detail]{2022ApJ...934...88T, 2023MNRAS.518.3326L}.
  The dust growth to several hundred microns in size can occur even in young disks. 
  The growth timescale of dust grains in the disks is estimated to be the order of $10^3$ yr and
  is much shorter than the age of Class 0/I young stellar objects (YSOs) \citep[see ][for more detailed estimate]{2022ApJ...934...88T}.
  Recent simulations actually show that dust growth and associated resistivity change can proceed in very young disks 
  \citep[or even in the first core;][]{2021ApJ...920L..35T,2023A&A...670A..61M, 2023PASJ...75..835T}.
  In addition, signs of dust growth to $\mm$ in size even in very young disks have been revealed by recent high-resolution observations
  \citep{2018NatAs...2..646H,2019ApJ...883...71C}.
  Therefore, there are theoretical and observational motivations
  for using equation (\ref{model_etaA}).
  }

$C$ is given as 
\begin{align}
C=\sqrt{\frac{m_i^2 \zeta}{m_g \beta_r}},
\end{align}
where $m_i$ and $m_g$ are the mass of ion and neutral particles.
{ In this paper, We assume that the major ion is HCO$^+$ and $m_i=29 m_p$ where $m_p$ is the proton mass.}
We also assume $m_g=2.34 m_p$.
$\zeta$ is the ionization rate. 

$\beta_r$ is the recombination rate and assumed to be
\begin{align}
 \label{beta_UMIST}
\beta_r=\beta_{r,0} \left(\frac{T}{300 ~{\rm K}}\right)^{-1} 
\end{align}
where $\beta_{r,0}=1.1 \times 10^{-7} ~{\rm cm^3 s^{-1}}$  which is { the rate of
HCO$^+$ and} taken from UMIST database \citep{2013A&A...550A..36M}.
Note that we adopted older value from RATE99  by \citet{1984ApJ...284L..13S}
to make power law simpler (latest recombination rate (RATE12) of UMIST gives $\beta_r \propto T^{-0.69}$).

$\gamma$ is given as
\begin{align}
\gamma=\frac{\langle \sigma v \rangle_{in}}{(m_g+m_i)},
\end{align}
where $\langle \sigma v \rangle_{in}$ is the rate coefficient for collisional momentum transfer between ions and neutrals.
We assume  $\langle \sigma v \rangle_{in} = 1.3 \times 10^{-9} ~{\rm cm^3 s^{-1}}$ which is calculated from the Langevin rate \citep{2008A&A...484...17P}.

For temperature, we assume the simple power law,
\begin{align}
  \label{model_temp}
  T(r)=T_0 \left(\frac{r}{\rm AU}\right)^{-1/2},
\end{align}
and vertically isothermal { just for simplicity.
However, it is straightforward to use a more complex temperature structures.}

We assume the sound velocity to be,
\begin{align}
  c_{\rm s}=1.9 \times 10^4 \left(\frac{T}{10 {\rm K}}\right)^{1/2} \cms.
\end{align}

\subsection{Assumptions of the model}
\label{AssumptionsOfModel}
In our governing equations, we (implicitly) assume that the internal angular momentum transport
mechanisms such as magneto-rotational instability (MRI), gravitational instability (GI) are negligible,
and magnetic braking is the mechanism which determines angular momentum evolution of the disk.
The justification for ignoring MRI and GI will be discussed after the actual disk structure is obtained.

Another important assumption is that the toroidal electric current is determined by radial gradient of $B_z$ as
\begin{align}
  J_\phi&=\frac{4 \pi}{c} (\nabla \times  \magB)_\phi= \frac{4 \pi}{c} (\frac{\partial B_r}{\partial z}- \frac{\partial B_z}{\partial r}) \nonumber \\
  &\sim  -\frac{4 \pi}{c} \frac{\partial B_z}{\partial r}.
\end{align}
This assumption is adopted in e.g., \citet[][]{1994ApJ...432..720B,2010A&A...521L..56D,2012A&A...541A..35D}
and leads to right hand side term of equation (\ref{eq5}).

On the other hand, another possibility is
\begin{align}
  J_\phi&\sim  \frac{4 \pi}{c} \frac{\partial B_r}{\partial z}.
\end{align}

If we further assume $B_r\sim B_z$ at $z=H$, the equation (\ref{eq5}) is replaced as
\begin{align}
  \label{eq5alt}
  B_z v_r=-\frac{\eta_A}{H} B_z.
\end{align}
We also investigated the solutions with this equation and found a few times weaker vertical magnetic field at $10$ AU
and slightly shallower profile of $B_z \propto r^{-29/24}$ because of the stronger radial diffusion for given magnetic field.
On the other hand, the density and radial velocity do not depend on the choice of
the azimuthal electric current.
We do not use the equation (\ref{eq5alt}) in this paper because the resulting magnetic field strength does not reproduce the simulation results in TMI23.

We assume that radial velocity for the magnetic flux advection
($v_r$ in the left hand side term of equation (\ref{eq5})) equals to the (vertically density weighted)
radial gas velocity.
We acknowledge that some previous studies show that the radial velocity at several scale height is much faster than the radial velocity at the midplane
\citep[so called layered accretion; e.g.,][]{1996ApJ...457..355G,2003ApJ...585..908F,2018ApJ...857....4T, 2018ApJ...857...34Z}.
And there are the discussions that $v_r$  in the equation (\ref{eq5}) (magnetic field transport velocity)
should be different from gas radial velocity \citep{2012MNRAS.424.2097G,2014MNRAS.441..852G}.
On the other hand, in the previous simulation studies started from cloud cores \citep{2014ApJ...796L..17M,2019ApJ...876..149M},
we do not observe such a significant inward accretion at several scale height rather the gas is significantly outflowing from several scale height.
We speculate that the difference of flow pattern at several scale height
is due to the significant difference of the magnetic field structure/strength in and around the disk among the studies,
and/or whether non-ideal MHD effects are considered. For example, the layered accretion becomes less pronounced or disappears
when ambipolar diffusion are taken into account \citep{2015ApJ...801...84G,2017ApJ...845...75B}.

The causes of the difference of flow pattern should be investigated in future studies.
At the same time, we think the validity of the assumptions made in a model should ultimately be verified observationally.
In this study, we provide the quantitative predictions of the radial profiles, mass, radius of the disk predicted from our model.
This will help to validate our disk model by future observational studies.

\subsection{Disk profiles}
\label{disk_structure}
By solving the equations (\ref{eq1}) to (\ref{eq6}) with the model for $\eta_A$ (equation (\ref{model_etaA})) and
temperature (equation (\ref{model_temp})) assuming the physical quantities have power law $f(r) \propto r^q$,
we can derive the physical quantities of disk.

{
  From equation (\ref{eq5}) and (\ref{eq6}), $v_r$ and  $B_{\phi, s}$ can be written as
  \begin{align}
    v_r&=-\frac{\eta_A}{r}, \nonumber \\ 
  B_{\phi, s} &=\left( \frac{H^3}{\eta_A r^2}\right) B_z v_\phi. \nonumber
  \end{align}
  By substituting these expressions into equations (\ref{eq1}) and  (\ref{eq4}),
  we obtain
  \begin{align}
    2 \pi \eta_A \Sigma &=\dot{M}_{\rm disk}, \nonumber\\
    H^3 B_z^2 &= \pi  \eta_A^2 \Sigma. \nonumber \\ 
  \end{align}
  Using equations (\ref{eq2}), (\ref{eq3}), (\ref{model_etaA}), (\ref{model_temp}), and $\Sigma=\sqrt{2 \pi} H \rho_{\rm mid} $
  these equations are the simultaneous equations for midplane density and vertical magnetic field.
  By solving the simultaneous equations and using the solutions for $\rho_{\rm mid}$ and $B_z$,
  }
the disk radial profiles are given as,
\begin{align}
  \label{rho}
  \rho_{\rm mid}(r)&=1.6\times 10^{-12} \nonumber \\
  &\mdotDisk^{\frac{2}{3}} \tempDisk^{-\frac{4}{3}} \nonumber \\
  &\massDisk \zetaCRDisk^{-\frac{1}{3}} \nonumber \\
  & \radDisk^{-\frac{7}{3}} \gcm,\\
    \label{Bz}
  B_z(r)&= 6.2\times 10^{-2} \nonumber \\
  &\mdotDisk^{\frac{2}{3}} \tempDisk^{-\frac{1}{3}} \nonumber \\
  &\massDisk^{\frac{1}{2}} \zetaCRDisk^{\frac{1}{6}} \nonumber \\
  & \radDisk^{-\frac{4}{3}} {\rm G},\\
    \label{vr}
  v_{r}(r)&= - 1.4 \times 10^3  \nonumber \\
  &\mdotDisk^{\frac{1}{3}} \tempDisk^{\frac{5}{6}} \nonumber \\
  & \massDisk^{-\frac{1}{2}} \zetaCRDisk^{\frac{1}{3}}  \nonumber \\
  & \radDisk^{\frac{1}{12}} \cms,\\
    \label{Bphi}
  B_{\phi}(r,z)&= -1.2\times 10^{-2}  \left(\frac{z}{H}\right) \nonumber \\
  &\mdotDisk^{\frac{1}{3}} \tempDisk^{\frac{1}{3}} \nonumber \\
  &\zetaCRDisk^{-\frac{1}{6}}  \nonumber \\
  & \radDisk^{-\frac{7}{6}} {\rm G},\\
    \label{etaA}
  \eta_A(r)&= 2.1\times 10^{17} \nonumber \\
  &\mdotDisk^{\frac{1}{3}} \tempDisk^{\frac{5}{6}} \nonumber \\
  &\massDisk^{-\frac{1}{2}} \zetaCRDisk^{\frac{1}{3}}  \nonumber \\
  & \radDisk^{\frac{13}{12}} {\rm cm^2 s^{-1}},
\end{align}
where we assume that $B_{\phi}$ is linear function of  $z/H$.

Then, the surface density is given as follows,
\begin{align}
    \label{sigma}
  \Sigma(r)&=\sqrt{2 \pi} H \rho_{\rm mid}=4.8\times 10^{1} \nonumber \\
  &\mdotDisk^{\frac{2}{3}} \tempDisk^{-\frac{5}{6}} \nonumber \\
  &\massDisk^{\frac{1}{2}} \zetaCRDisk^{-\frac{1}{3}}  \nonumber \\
  &\radDisk^{-\frac{13}{12}} \gcmcm.
\end{align}

Plasma $\beta$ and "Elsasser number" for ambipolar diffusion Am at midplane are given as
\begin{align}
  \label{beta}
  \beta(r) &= \frac{P_{\rm gas}}{P_{\rm mag}} =3.0 \times 10^1 \nonumber \\
  &\mdotDisk^{-\frac{2}{3}} \tempDisk^{\frac{1}{3}} \nonumber \\
  &\zetaCRDisk^{-\frac{2}{3}} \radDisk^{-\frac{1}{6}},
\end{align}
and 
\begin{align}
  \label{Am}
  {\rm Am}(r) &=\frac{v_A^2}{\eta_A \Omega}=2.7 \times 10^{-1} \nonumber \\
  &\mdotDisk^{\frac{1}{3}} \tempDisk^{-\frac{1}{6}} \nonumber \\
  &\zetaCRDisk^{\frac{1}{3}} \radDisk^{\frac{1}{12}}.
\end{align}
{ Here, $P_{\rm gas}=\gamma \rho_{\rm mid} c_s^2$ and $P_{\rm mag}=B_z^2/(8 \pi)$ are the gas and magnetic pressure, respectively (where $\gamma=5/3$ is specific heat ratio).
$v_A=B_z/\sqrt{4 \pi \rho_{\rm mid}}$ is the Alfv\'{e}n velocity.}

Plasma $\beta$ and Am have a simple relation of,
\begin{align}
  \label{beta_Am}
  \beta({\rm Am})&=2.1 {\rm Am}^{-2}.
\end{align}

\begin{figure}
  \includegraphics[trim=0mm 0mm 0mm 0mm,width=70mm,angle=0]{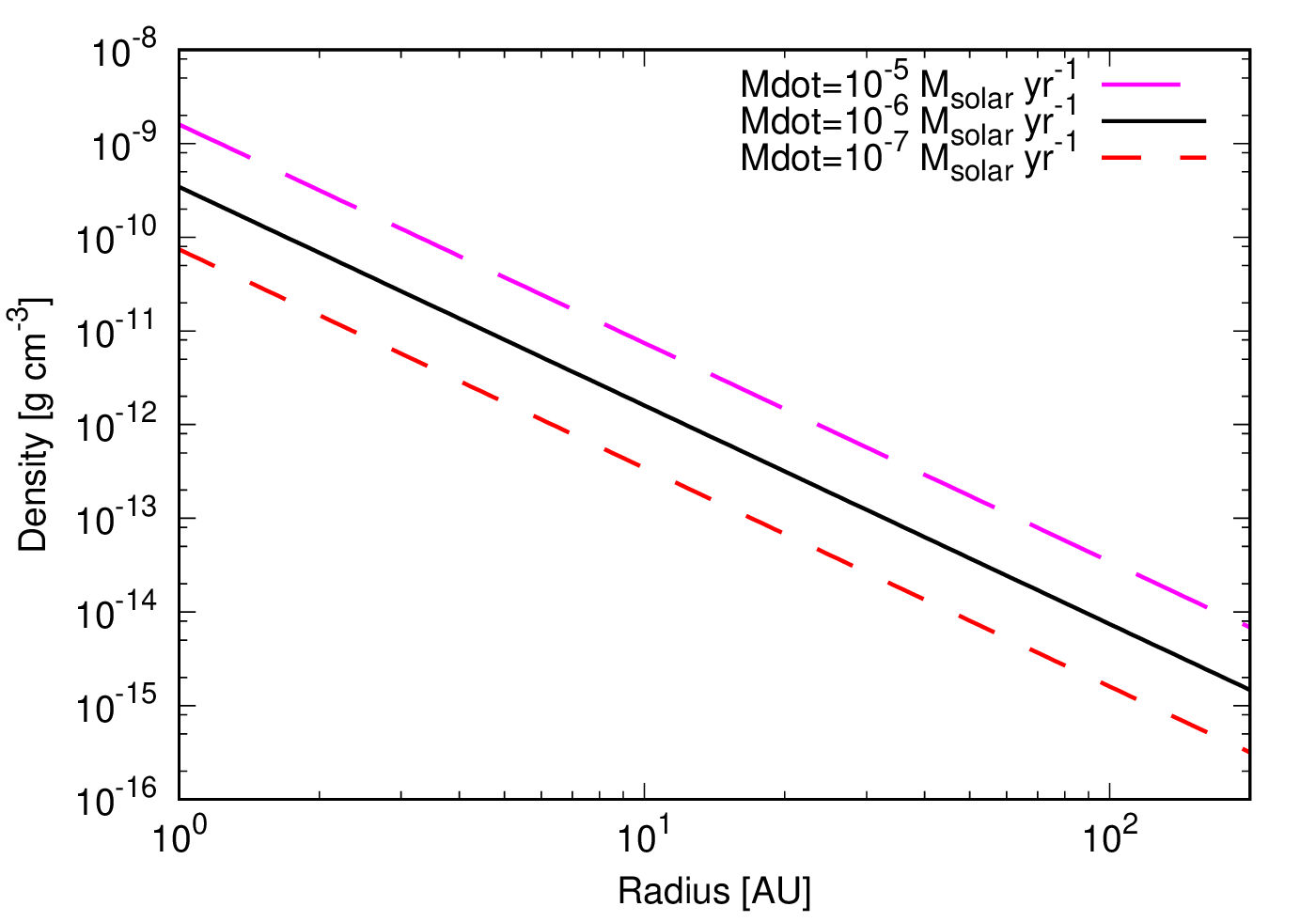}
  \includegraphics[trim=0mm 0mm 0mm 0mm,width=70mm,angle=0]{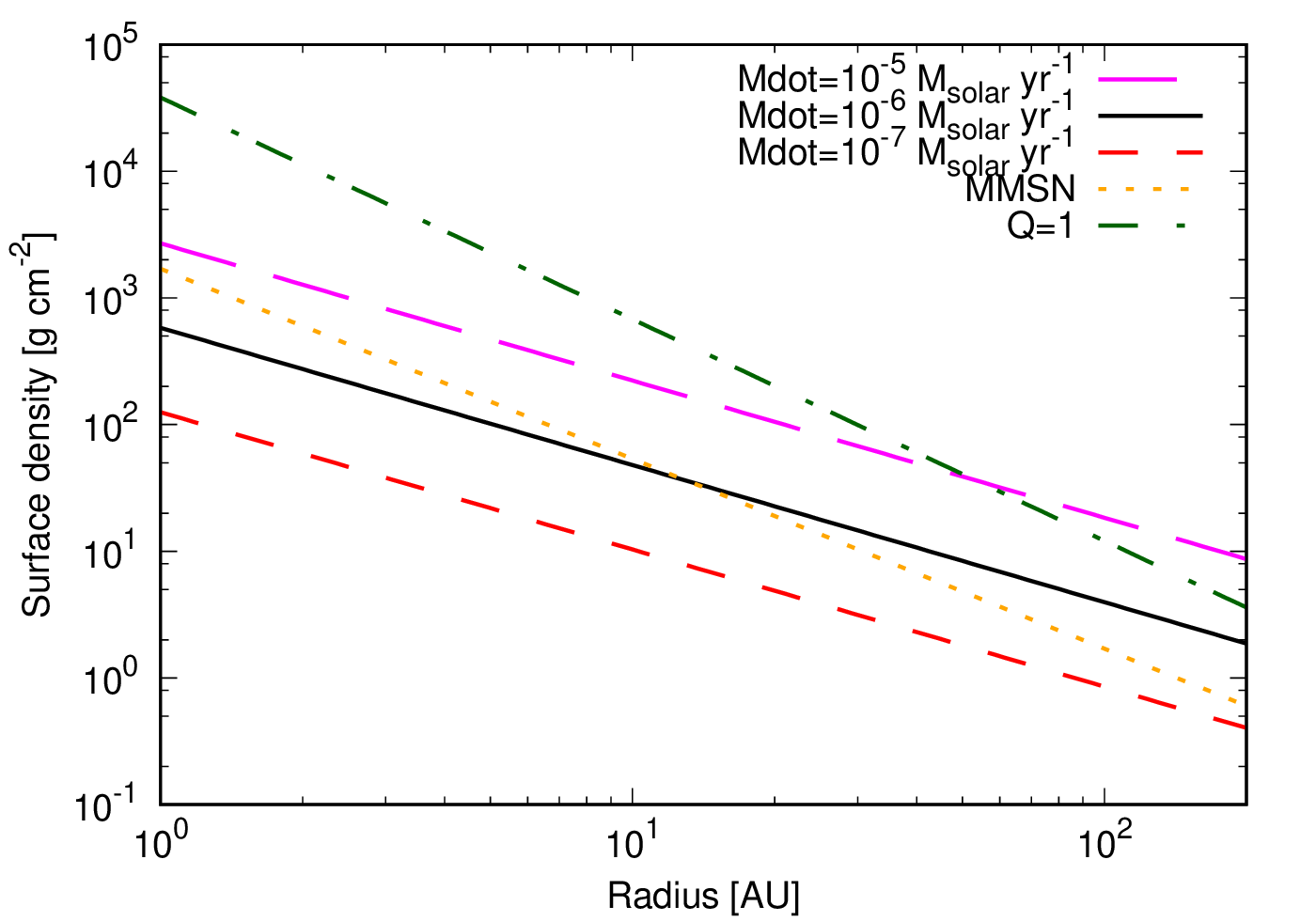}
  \caption{
    Radial profiles of midplane density and surface density.
    The parameters of $M_{\rm star}=0.3 \msun$, $T_0=150$ K, $\zeta=10^{-18} {\rm ~s^{-1}}$ are used.
    Long dashed (magenta), solid (black), and short dashed (red) lines show the profile with
    $\dot{M}_{\rm disk}=10^{-5}, 10^{-6}, 10^{-7} \msunyr$, respectively.
    dotted-dashed (green) and dotted (orange) lines show the surface density of gravitationally unstable disk
    and the Minimum Mass Solar Nebula (MMSN) model ($\Sigma_{\rm MMSN}=1.7\times 10^3 (r/{\rm AU}) \gcmcm$) \citep{1981PThPS..70...35H}, respectively.
}
\label{r_rho}
\end{figure}

Figure \ref{r_rho} shows the radial profile of midplane density (equation (\ref{rho})) and surface density of the model (equation (\ref{sigma})).
It shows that the surface density is more or less consistent with the the Minimum Mass Solar Nebula (MMSN) model \citep{1981PThPS..70...35H}
when $\dot{M}_{\rm disk} \sim 10^{-6} \msunyr$.
We also plot the surface density of gravitationally unstable disk $\Sigma_{\rm GI}$
in the figure calculated from the condition that Toomre's $Q$ value is $Q=1$
and the temperature profile of equation (\ref{model_temp}),
\begin{align}
    \label{sigmaGI}
  \Sigma_{\rm GI}(r)&=\frac{c_s \Omega}{\pi G Q}=6.8\times 10^{2} \nonumber \\
  &\tempDisk^{\frac{1}{2}} \massDisk^{\frac{1}{2}}   \nonumber \\
  &\radDisk^{-\frac{7}{4}} \gcmcm.
\end{align}
With large mass accretion rate of
$\dot{M}_{\rm disk} \sim 10^{-5} \msunyr$ which is almost upper limit of mass accretion rate for low-mass protostar formation,
the surface density becomes $\Sigma_{\rm GI}<\Sigma$ in $r\gtrsim 100$ AU, and gravitational instability may develop there
even with magnetic braking.
On the other hand, once mass accretion rate becomes $\dot{M}_{\rm disk} \lesssim 10^{-6} \msunyr$, the surface density satisfies
$\Sigma_{\rm GI}>\Sigma$ in r$<100$ AU.
Although we will neglect gravitational instability for our disk model in the rest sections of this paper,
it is important to recognize that gravitational instability can develop if the disk becomes larger than $\sim 100$ AU
at a stage where the mass accretion rate is $\dot{M}_{\rm disk} \sim 10^{-5} \msunyr$.

\begin{figure}
  \includegraphics[trim=0mm 0mm 0mm 0mm,width=70mm,angle=0]{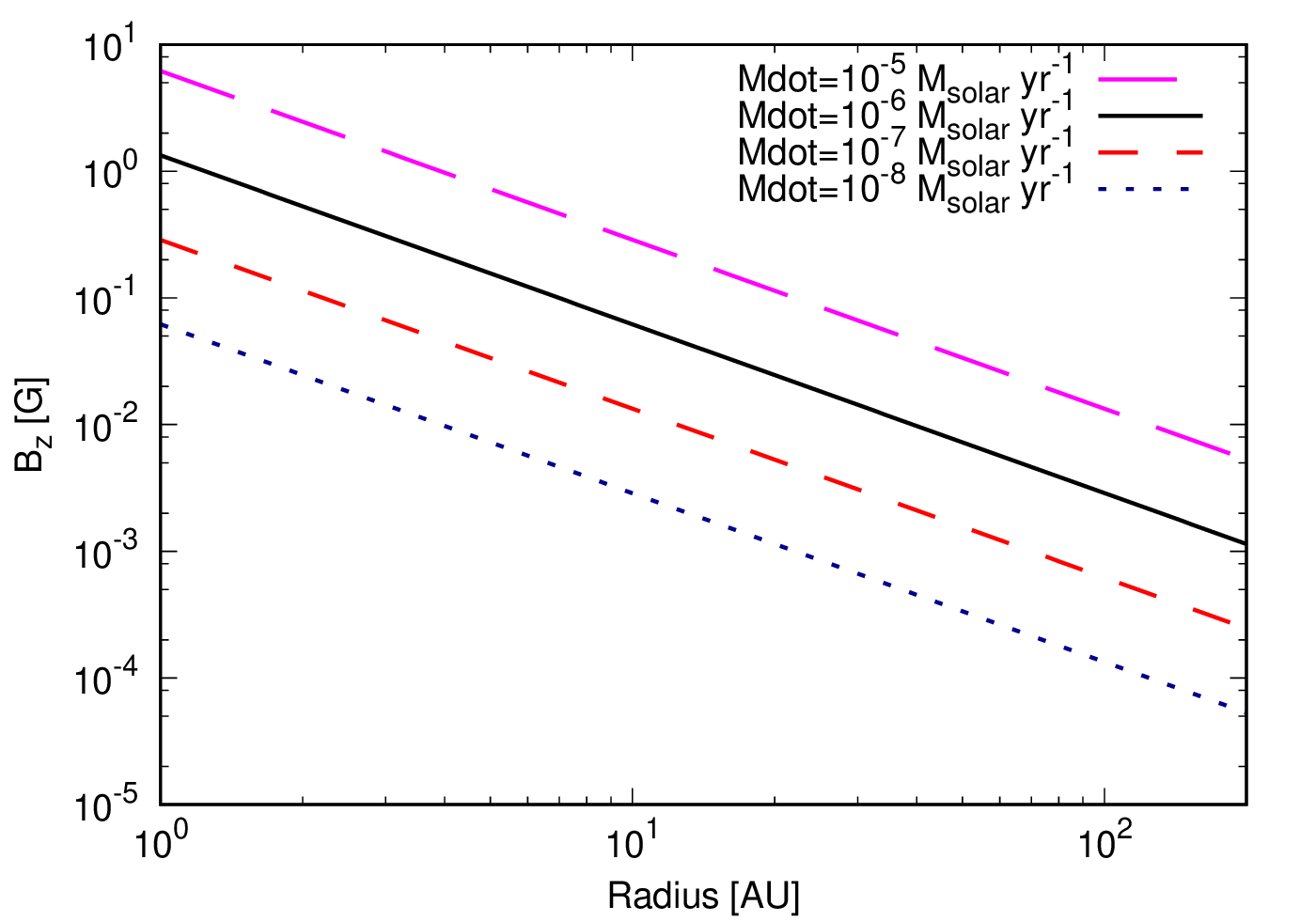}
  \includegraphics[trim=0mm 0mm 0mm 0mm,width=70mm,angle=0]{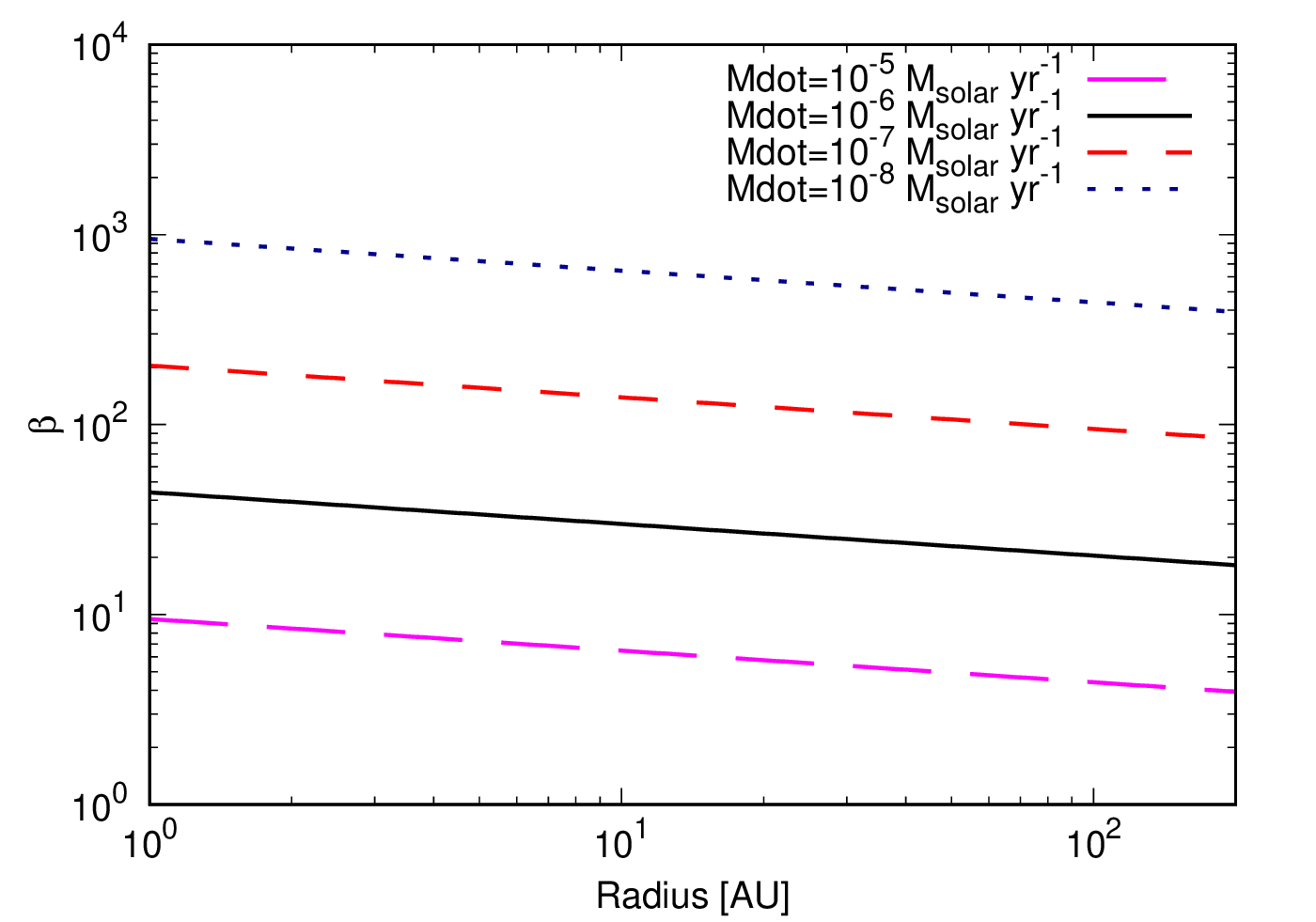}
  \caption{
    Radial profiles of vertical magnetic field and plasma $\beta$.
    The parameters of $M_{\rm star}=0.3 \msun$, $T_0=150$ K, $\zeta=10^{-18} {\rm ~s^{-1}}$ are used.
    Long dashed (magenta), solid (black), short dashed (red), and dotted (blue) lines show the profile with  $\dot{M}_{\rm disk}=10^{-5}, 10^{-6},  10^{-7},  10^{-8} \msunyr$, respectively.
}
\label{r_Bz}
\end{figure}

Figure \ref{r_Bz} shows the radial profiles of vertical magnetic field (equation (\ref{Bz})) and plasma $\beta$ (equation (\ref{beta})).
Our model predict that, { as the protostar evolves and the mass accretion rate decreases},
the vertical magnetic field decreases as $\propto \dot{M}^{2/3}_{\rm disk}$ and midplane plasma $\beta$ increases
as $\propto \dot{M}^{-2/3}_{\rm disk}$.
With the $\dot{M}_{\rm disk} \sim 10^{-6} \msunyr$ which is the typical value of the Class 0/I YSOs,
the vertical magnetic field and $\beta$ at $10$ AU are $B_z\sim 60 ~{\rm mG}$  and $\beta\sim 30$, respectively.
This indicates that the disk needs this level of the magnetic field strength to cause
the sufficient mass accretion of $\dot{M}_{\rm disk} \sim 10^{-6} \msunyr$ only by magnetic braking
without resorting to other mass accretion mechanisms such as gravitational instability.
Note that the magnetic field strength are consistent with the disk formation simulations starting from collapsing cloud cores \citep[e.g.,][]{2016A&A...587A..32M}.

On the other hand, with the $\dot{M}_{\rm disk} \sim 10^{-8} \msunyr$ which may be suitable for late Class I YSOs to Class II objects,
the vertical magnetic field and $\beta$ at $10$ AU are $B_z\sim 3 ~{\rm mG}$  and $\beta\sim 650$, respectively.
This magnetic field profile may be suitable for the initial conditions of simulations of isolated star-disk systems just after the envelope is depleted.

Note that our disk model predicts a stronger magnetic field than the typical initial magnetic field strength of the simulations that examine the impact of magnetic field
in isolated star-disk systems \citep[typically $\beta>10^3$;][]{2014ApJ...784..121S,2015ApJ...801...84G,2017ApJ...845...75B,2018ApJ...857...34Z}.
By examining the new parameter space predicted from our disk model with future high-resolution simulations, a new disk evolution picture is expected to emerge.

\begin{figure}
  \includegraphics[trim=0mm 0mm 0mm 0mm,width=70mm,angle=0]{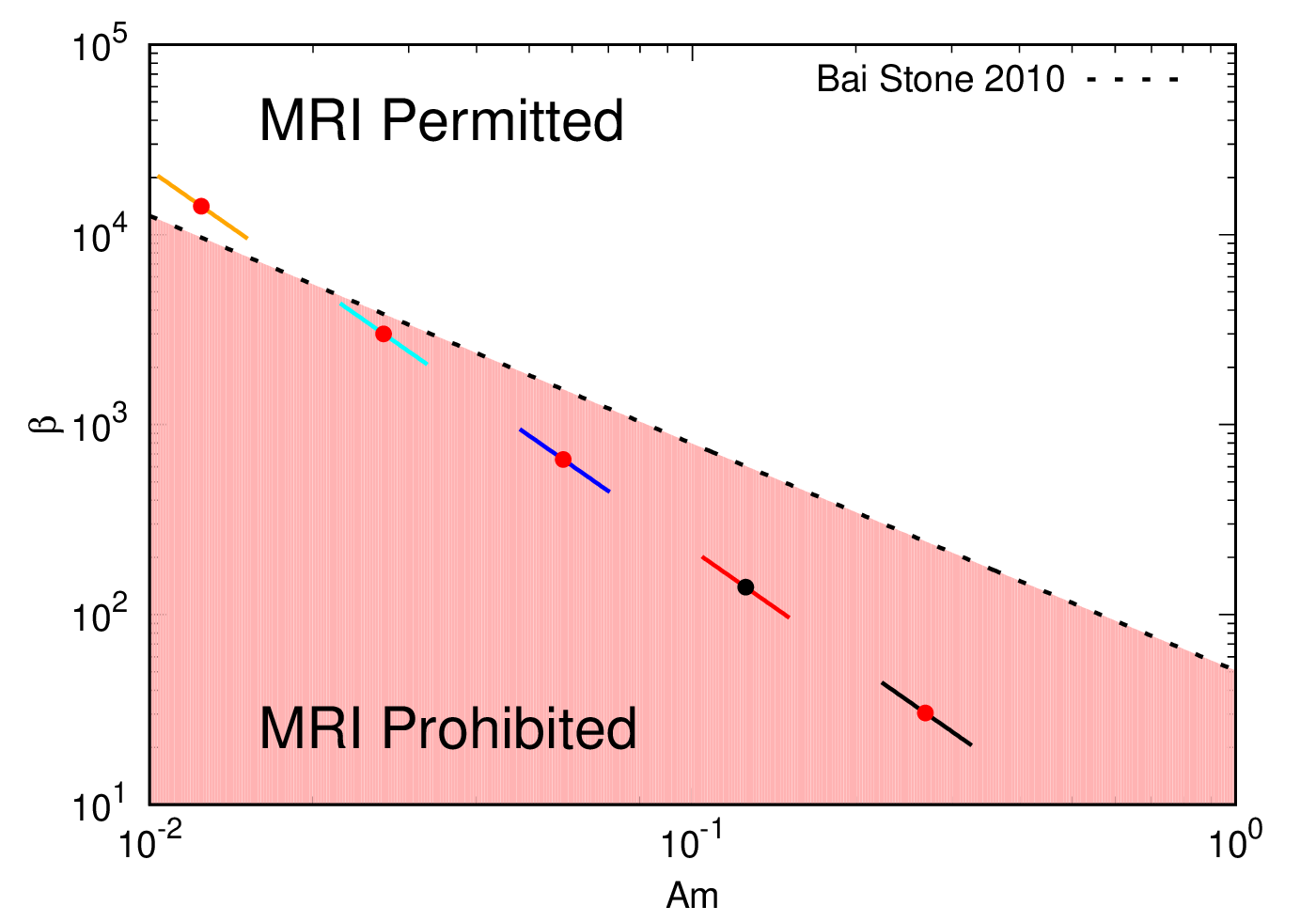}
    \caption{
    Plasma $\beta$ of the disk as a function of Am.
    The parameters of $M_{\rm star}=0.3 \msun$, $T_0=150$ K, $\zeta=10^{-18} {\rm ~s^{-1}}$ are used.
    black, red, blue, and cyan 
    lines show the profile of the disk with  $\dot{M}_{\rm disk}=10^{-6},  10^{-7}, 10^{-8}, 10^{-9}$ and $10^{-10} \msunyr$, respectively.
    Upper left end of each line  corresponds to $r=1$ AU and Lower right end of each line corresponds to $r=100$ AU.
    The red (or black) point on each line corresponds to $r=10$ AU.
    The dashed line shows the boundary between whether MRI operates or not,
    and in the region below this line, hatched in light red, MRI  does not operate
    \citep{2011ApJ...736..144B}.
}
\label{Am_beta}

\end{figure}

\citet{2011ApJ...736..144B} { performed three-dimensional shearing-box simulations over a wide range of parameter space,
and revealed the parameter space on $\beta$-${\rm Am}$ plane in which magnetorotational instability (MRI) operates,
corresponding to the requirement that the most unstable vertical wavelength of MRI should be less than the disk scale height.}
With the aid of their conditions, we can discuss whether MRI develops in our disk model, and whether it is reasonable to neglect MRI.

Figure \ref{Am_beta} shows plasma  $\beta$ (equation (\ref{beta})) of the disk as a function of Am (equation (\ref{Am})).
In this plot, we consider the disk radius of $1 ~{\rm AU} < r< 100 ~{\rm AU}$. We derived radius of given Am from equation (\ref{Am}) and
calculate the plasma $\beta$ at the obtained radius. Upper left end of each line corresponds to $r=1$ AU, the symbol on the each line corresponds to $r=10$ AU,
and lower right end of each line corresponds to $r=100$ AU.
The different lines represent the different mass accretion rate from $10^{-6} \msunyr $ to $ 10^{-10} \msunyr$.
The figure shows that with our fiducial parameters ($M_{\rm star}=0.3 \msun$, $T_0=150$ K, $\zeta=10^{-18} {\rm ~s^{-1}}$),
{ MRI becomes active in $r \lesssim 100$ AU when the mass accretion rate satisfies $\dot{M}_{\rm disk} \lesssim 10^{-10} \msunyr$.}
In this paper, we are interested in Class 0/I YSOs and the mass accretion rate in this phase is expected to be larger than $10^{-9} \msunyr$
and neglecting MRI in our disk model is justified.

This figure indicates that MRI is difficult to develop during the early evolutionary stage of the disk
such as those with $\dot{M}_{\rm disk} \gtrsim 10^{-9} \msunyr$.
For MRI to develop, Am and $\beta$ must be simultaneously large.
However, when Am is large, the outward transport of magnetic flux by ambipolar
diffusion becomes inefficient and the vertical magnetic field of the disk increases due to the radial advection, and hence $\beta$ decreases.
On the other hand, when ambipolar diffusion works efficiently, the vertical magnetic field is extracted by outward diffusion and $\beta$ increases.
Because of this self-regulative behavior of radial advection and outward diffusion of vertical magnetic field,
the disk evolves towards the upper left direction on the $\beta$-Am plane as the mass accretion rate decreases.

The possible scenario in which MRI develops even with larger $\dot{M}_{\rm disk}$ is the case that cosmic ray are strongly shielded
and the ionization rate in the disk is lowered.
Since Am and $\beta$ depend on the mass accretion rate and the ionization rate as ${\rm Am}\propto (\dot{M}_{\rm disk} \zeta)^{1/3}$
and $\beta \propto (\dot{M}_{\rm disk} \zeta)^{-2/3}$, respectively,
decreasing the ionization rate by one order of magnitude and decreasing the mass accretion rate by one order of magnitude
have the same impact for Am and $\beta$. { Thus, for example, if $\zeta$ decreases to $10^{-22}  {\rm ~s^{-1}}$
in a disk due to low cosmic-ray ionization rate and lack of short-lived radionuclides,
MRI can develop in the disk even when $\dot{M}_{\rm disk} \sim 10^{-6} \msunyr$
because the orange line can also be regarded as a disk with $\dot{M}_{\rm disk}=10^{-6} \msunyr$ and $\zeta=10^{-22} {\rm ~s^{-1}}$.}
In this case, outer disk would be gravitationally unstable because $\Sigma \propto \zeta^{-1/3}$
and it may become $\Sigma \gtrsim \Sigma_{\rm GI}$ in $r\gtrsim$ 10 AU (see figure \ref{r_rho})
The another possible scenario in which MRI develops is the case the electric current is given as $J_\phi \sim B_z/H$ { (as discussed in \S \ref{AssumptionsOfModel})}.
In this case,  the magnetic field becomes a few times smaller and plasma $\beta$ increases about one order of magnitude, and disk can enter the MRI permitted region.

\subsection{Mass, magnetic flux, and mass-to-flux ratio of disk}
By radially integrating $\Sigma(r)$ and $B_z(r)$, we can obtain the disk mass, disk magnetic flux and mass-to-flux ratio of the protostar-disk system.
\begin{align}
  \label{Mdisk}
  M_{\rm disk}&=\int_0^{r_{\rm disk}} 2 \pi r \Sigma(r) d r=3.0 \times 10^{-2} \nonumber \\
  &\mdotDisk^{\frac{2}{3}} \tempDisk^{-\frac{5}{6}} \nonumber \\
  &\massDisk^{\frac{1}{2}} \zetaCRDisk^{-\frac{1}{3}}  \nonumber \\
  &\radDiskSize^{\frac{11}{12}} \msun,
\end{align}
where $r_{\rm disk}$ is the disk radius.
Note that the radial power index of $\Sigma$ is larger than $-2$ and lower limit
of integral does not contribute to the disk mass and we set it to be zero.

the magnetic flux of the disk is calculated as
\begin{align}
  \label{Phi_mag}
  \Phi_{\rm disk}&=\int_0^{r_{\rm disk}} 2 \pi r B_z(r) d r=2.7 \times 10^2 \nonumber \\
  &\mdotDisk^{\frac{2}{3}} \tempDisk^{-\frac{1}{3}} \nonumber \\
  &\massDisk^{\frac{1}{2}} \zetaCRDisk^{\frac{1}{6}}  \nonumber \\
  &\radDiskSize^{\frac{2}{3}} \gauau.
\end{align}

By assuming that the protostar has the most of the mass and the disk has the most of the magnetic flux of the system,
the mass-to-flux ratio of the system can be estimated as,
\begin{align}
  \label{mu_mag}
\mu &\equiv  \left(\frac{M_{\rm star}}{\Phi_{\rm disk}}\right)/\left(\frac{M}{\Phi}\right)_{\rm crit}= 2.0 \times 10^1  \nonumber \\
  & \mdotDisk^{-\frac{2}{3}} \tempDisk^{\frac{1}{3}} \nonumber \\
  &\massDisk^{\frac{1}{2}} \zetaCRDisk^{-\frac{1}{6}} \radDiskSize^{-\frac{2}{3}},
\end{align}
where we assume $\left(\frac{M}{\Phi}\right)_{\rm crit}=\frac{0.53}{3 \pi}\left(\frac{5}{G}\right)^{1/2}$ \citep{1976ApJ...210..326M}.

\begin{figure}
  \includegraphics[trim=0mm 0mm 0mm 0mm,width=70mm,angle=0]{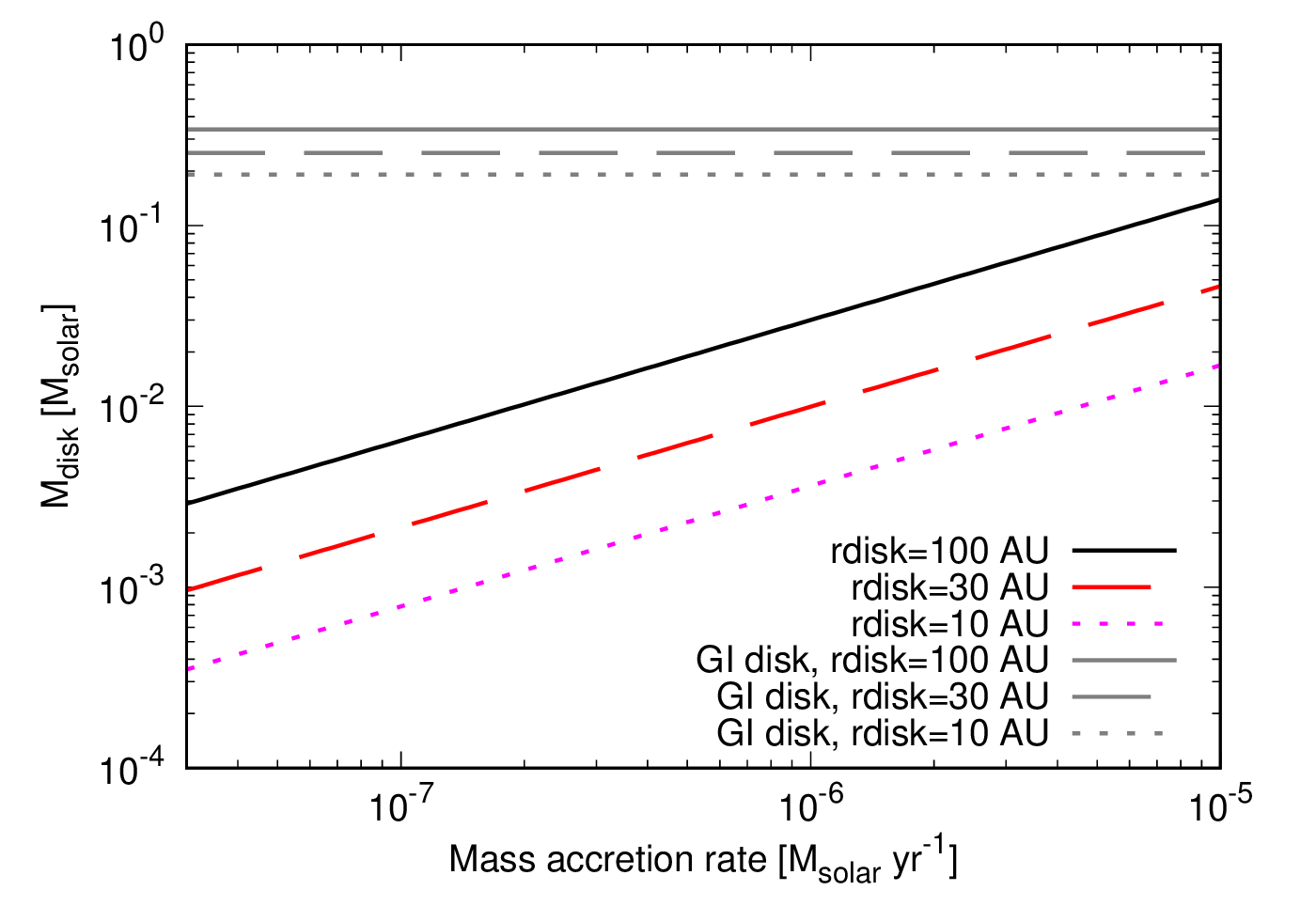}
  \includegraphics[trim=0mm 0mm 0mm 0mm,width=70mm,angle=0]{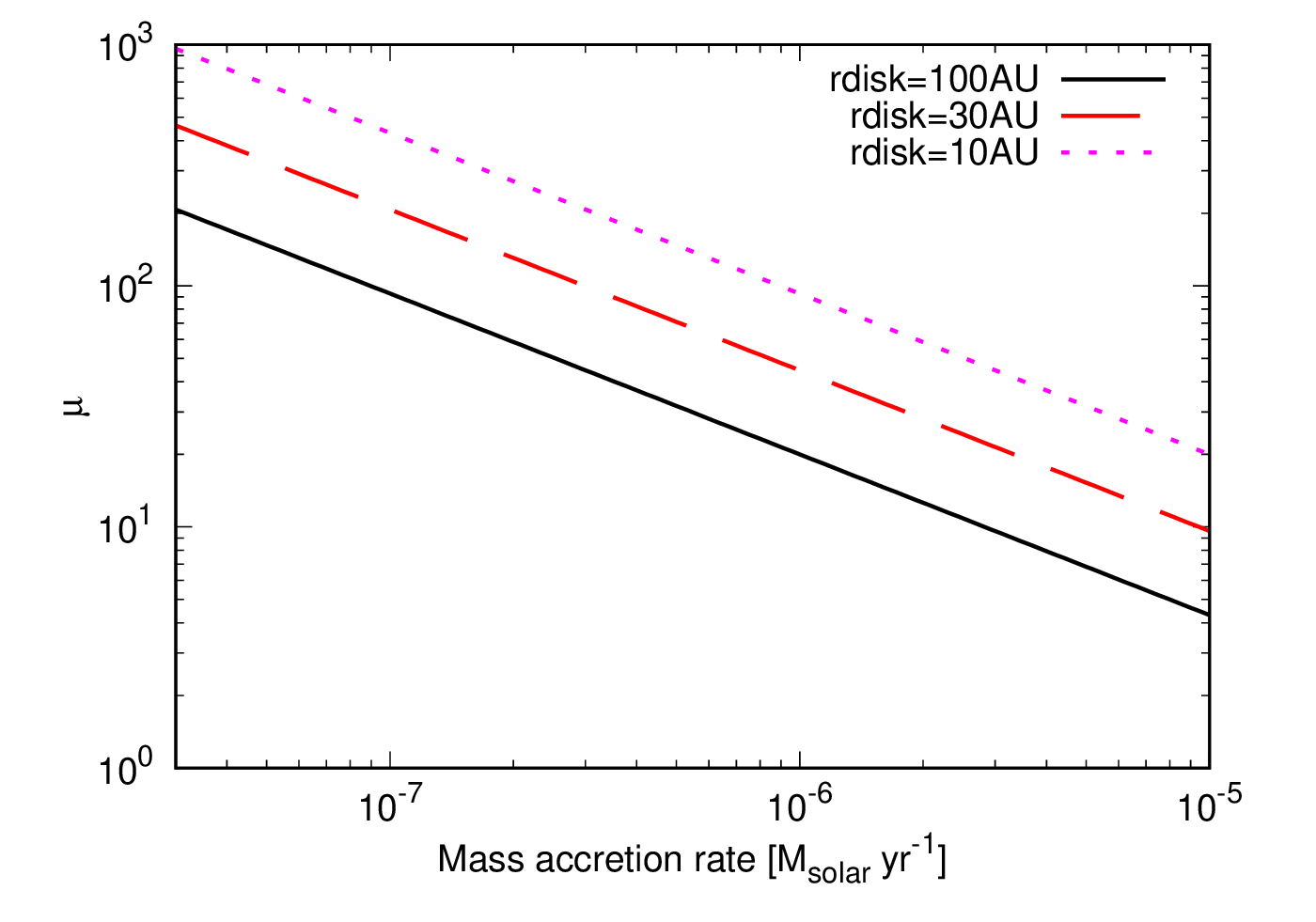}
      \caption{
    Disk mass and mass-to-flux ratio as a function of mass accretion rate.
    The parameters of $M_{\rm star}=0.3 \msun$, $T_0=150$ K, $\zeta=10^{-18} {\rm ~s^{-1}}$ are used.
    Solid (black), dashed (red), and dotted (magenta)
    lines show the profile with  $r_{\rm disk}=100, 30,  10$ AU, respectively.
    { The gray horizontal lines in the disk mass show the mass of GI disks (equation (\ref{Mdisk})).}
}
\label{Mdot_Mdisk}

\end{figure}

Figure \ref{Mdot_Mdisk} shows the disk mass and mass-to-flux ratio $\mu$ as a function of mass accretion rate.
The disk mass is $0.01 \msun \lesssim M \lesssim 0.1 \msun$ for mass accretion rates $\dot{M}_{\rm disk} \gtrsim 10^{-6} \msunyr$ and several 10 AU sized disks.
This value is consistent to the disk mass suggested from observations \citep{2020ApJ...890..130T,2020A&A...640A..19T}.
{ 
In the figure, we also plot the mass of gravitationally unstable disk,
\begin{align}
  \label{Mdisk}
  M_{\rm GI, disk}&=\int_0^{r_{\rm disk}} 2 \pi r \Sigma_{\rm GI}(r) d r=3.4 \times 10^{-1} \nonumber \\
  &\tempDisk^{\frac{1}{2}} \massDisk^{\frac{1}{2}}  \nonumber \\
  &\radDiskSize^{\frac{1}{4}} \msun.
\end{align}
The mass of gravitationally unstable disk is $ M_{\rm GI, disk} \gtrsim 0.1 \msun$.
This value seems not to be consistent to the disk mass estimated from observations
although the disk mass estimate from the observations also has uncertainties such as the uncertainty
of dust opacity or the optical thickness \citep[see][]{2017ApJ...838..151T,2023ASPC..534..317T}.

It is worth noting that the mass of gravitationally unstable disk depends weakly on the radius.
Since the disk size has a large dispersion \citep{2019A&A...621A..76M},
it may be possible to determine whether the disk is gravitationally unstable or not
by investigating whether the disk mass has a correspondingly large dispersion.
}

The mass-to-flux ratio is $\mu \gtrsim 10$
{ unless high mass accretion rate ($\dot{M}_{\rm disk} \sim 10^{-5} \msunyr$) and large disk $r \sim 100$ AU are realized simultaneously}.
This value is sufficiently larger than the mass-to-flux ratio of the molecular cloud core ($\mu \sim 1$).
In other words, the magnetic flux brought into the disk by mass accretion from envelope is sufficient
to realize $B_z$ shown in equation (\ref{Bz}). More precisely, the disk { with size of $r \lesssim 100$ AU }cannot hold the all magnetic flux of the cloud core in it { and the magnetic flux piles up around the disk which may cause the ion-neutral drift around the disk}
\citep[][]{2011ApJ...738..180L,2016A&A...587A..32M,2020ApJ...896..158T,2020ApJ...900..180M,2021MNRAS.505.5142Z}.
Although the vertical magnetic field strength of the disk given by equation (\ref{Bz})
may seem very large, it turns out to be very small compared to the magnetic flux possessed by the molecular cloud core. 

To connect the disk size with the angular momentum accretion rate from the envelope,
we estimate the angular momentum removal rate by the magnetic torque, which is calculated as,
\begin{align}
  \dot{J}_{\rm out}&=\int^{r_{\rm disk}}_{0} r \frac{B_{\phi, s} B_z}{4 \pi H}~ 2 H~ 2 \pi r dr=\dot{M}_{\rm disk}\sqrt{G M r_{\rm disk}} \nonumber \\
  &=1.5\times 10^{40} \mdotDisk \nonumber \\
  &\massDisk^{\frac{1}{2}}\radDiskSize^{\frac{1}{2}} {\rm g~ cm^2~ s^{-2}}.
\end{align}
This simple form of $\dot{J}_{\rm out}$ is immediately derived from equations  (\ref{eq1}) and (\ref{eq4}),
i.e., the angular momentum removal should balance with the angular momentum supply by radial advection at each radius.

If the disk evolves by the magnetic braking and is in a steady state, the angular momentum removal
from the disk should balance with the angular momentum supply from the envelope.
Thus, we can assume $\dot{J}_{\rm out}=\dot{J}_{\rm env}~( \equiv \dot{M}_{\rm env} j_{\rm env})$
where $\dot{M}_{\rm env}$ and $j_{\rm env}$ are the mass accretion rate from the envelope to disk and specific angular momentum of the accretion flow, respectively.
By further assuming $\dot{M}_{\rm disk}=\dot{M}_{\rm env}$ i.e., the  mass accretion rate from the envelope to disk equals to the mass accretion rate within the disk,
we can estimate the disk size as 
\begin{align}
  r_{\rm disk}=160 \jEnv^2 \massDisk^{-1} {\rm AU},
\end{align}
which is actually showing that the disk radius is equal to the centrifugal radius of the accretion flow,
\begin{align}
  \label{r_disk1}
  r_{\rm disk}=r_{\rm cent} \equiv \frac{j_{\rm env}^2}{G M_{\rm star}}.
\end{align}
This result may seem obvious at first glance.
Note, however, that this does not hold for disks that evolve
by internal angular momentum transport within them (e.g., viscous accretion due to MRI or GI).

If the disk evolves by internal viscous accretion,
the angular momentum is reserved in the disk and the radius of the disk is determined by the total angular momentum having brought into the disk.
Thus, the disk radius should depend on the entire accretion history, not on the instantaneous angular momentum accretion from the envelope.
In other words, $r_{\rm disk}$ should be calculated from the equation
\begin{align}
\int^{r_{\rm disk}}_0 \Sigma~ r^2 \Omega~ 2\pi r dr=\int^t_0 \dot{M}_{\rm env}(t) j_{\rm env}(t) dt.
\end{align}

In this case, the specific angular momentum at the outer edge of the disk is in general different from the specific angular momentum of the envelope accretion
at the envelope-to-disk boundary \citep[see, figure 2 of ][ for example]{2016MNRAS.463.1390T}.
This provides an observational test to determine whether the disk is evolving by magnetic braking or by internal angular momentum transport.
We discuss this point at the end of the next section.

\section{Disk evolution under the mass and angular momentum accretion from the envelope}
In this section, we discuss the time evolution of protoplanetary disks in Class 0/I phase by combining the disk model constructed
in the previous section with a model of an envelope accretion.

\begin{figure}
  \includegraphics[trim=0mm 0mm 0mm 0mm,width=80mm]{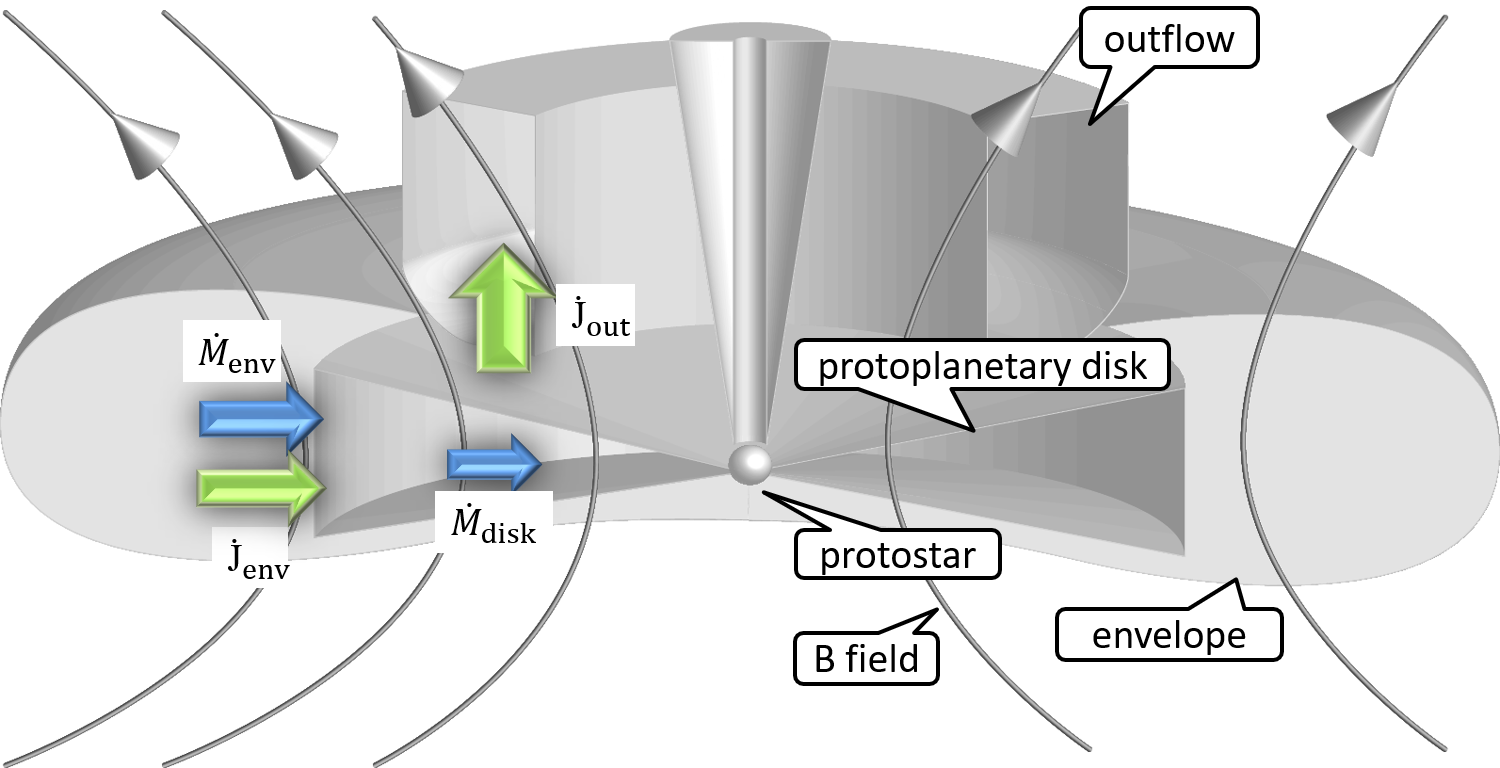}
  \caption{
    The schematic figure of the system we consider in this section.
}
\label{schematic}
\end{figure}

Figure \ref{schematic} shows the schematic picture of envelope-disk system considered in this section.
We assume that mass  and angular momentum accrete to the outer edge of the disk.
This is based on the observations that Class 0/I YSOs commonly have bipolar outflows in the normal direction of the disk,
making the accretion from the normal direction difficult.
Furthermore, we assume that the total mass of the disk and protostar is equal to the enclosed mass inside a certain radius of the collapsing envelope.
By relating the mass and angular momentum accretion, and enclosed mass of envelope models to the mass accretion rate in the disk,
the angular momentum extraction rate, and the mass of the central protostar and disk of our disk model,
we can investigate the time evolution of the disk with the mass accretion from the envelope.

\subsection{Envelope accretion model}
For the envelope accretion model, we consider the pressure free spherical symmetric collapse of \citet{2005MNRAS.360..675V}.

The equation of motion of spherical shell is
\begin{align}
\frac{d v_r}{d t}=-G\frac{M(r)}{r^2},
\end{align}
where $v_r$ and $r$ are the radial velocity and radius of the spherical shell.
$M(r)$ is the enclosed mass within the $r$.
By solving this equation with the assumption that $v_r=0$ at $t=0$, we have
\begin{align}
  \sqrt{2 G \frac{M(r_0)}{r_0^3}} t&=\arccos\left(\sqrt{\frac{r}{r_0}}\right)+\frac{1}{2}\sin\left(2\arccos \left(\sqrt{\frac{r}{r_0}}\right)\right),
\end{align}
where $r_0$ is the radius at $t=0$.
By solving this equation for given $t$ and $r$, we have $r_0(r,t)$.
Then, the radial velocity, density, and mass accretion, rate, and specific angular momentum can be obtained as,
\begin{align}
  v_r(r, t)&=\sqrt{2 G M(r_0)\left( \frac{1}{r}-\frac{1}{r_0} \right)},\\
  \rho(r,t)&=\rho_0(r_0)\frac{r_0^2}{r^2}\frac{d r_0}{d r},\\
  M(r, t)&=M(r_0),\\
  \dot{M}_{\rm env}(r, t) &=4 \pi r^2 \rho(r,t) v_r(r, t),\\
  j_{\rm env}(r, t)&=\epsilon_{\rm mag} f_J r_0^2 \Omega_0(r_0) .
\end{align}
where $\Omega_0(r_0)$ is the angular velocity at the equator of the spherical shell.
$f_J$ is the correction factor that arises when the angular momentum of the rotating spherical shell is described by the angular velocity at the equator of the spherical shell.
For example $f_J=2/3$ if the shell is rigidly rotating (note that the moment of inertia is $I=2/3 M r^2$ for the rigidly rotating shell),
and $f_J=0.72$ if $\Omega \propto r^{-1/2}$. For arbitrary angular velocities of $\Omega \propto r^{\alpha}$,  $ f_J=\sqrt{\pi} \Gamma(2+\alpha/2)/(2 \Gamma(5/2+\alpha/2))$ for $\alpha>-4$.
$\epsilon_{\rm mag}$ is a factor to mimic the magnetic braking in the envelope.
We assume $\epsilon_{\rm mag}=1/3$ just for simplicity.
Note that the early studies have suggest the  $\epsilon_{\rm mag} \sim 1/3$ \citep{1994ApJ...432..720B,2002ApJ...575..306T}; but it may change in the late accretion phase { or magnetic field geometry}
\citep[e.g.,][]{2012A&A...543A.128J,2020ApJ...900..180M}.
{ We believe that determining the value of $\epsilon_{\rm mag}$ precisely is one of the major issues for future research.}

Here,
\begin{align}
  \frac{d r}{d r_0}&=\frac{r}{r_0} \nonumber \\
  &-\sqrt{\frac{G}{2 M(r_0) r_0}} \left(\frac{d M(r_0)}{d r_0}-\frac{3 M(r_0)}{r_0}\right) \sin\left(2\arccos \sqrt{\frac{r}{r_0}}\right).
\end{align}

By specifying the initial density and rotation profile, we obtain $M(r,t), \dot{M}_{\rm env}(r, t), j_{\rm env}(r, t)$.
In this paper, we consider the Bonnor-Ebert sphere as the initial density profile,
\begin{eqnarray}
  \rho_0(r)= \varrho_0 \Psi_{\rm BE}(r/a) ~{\rm for},
\end{eqnarray}
and
\begin{eqnarray}
  a=c_{\rm s, iso} \left( \frac{1}{4 \pi G \varrho_0} \right)^{1/2}.
\end{eqnarray}
where $\Psi_{\rm BE}$ is non-dimensional density profile of the critical Bonnor-Ebert sphere,
$\varrho_0$ is the central density,
and $R_c=1.82 a$ is the characteristics radius of the Bonnor-Ebert sphere.

{
$\Psi_{\rm BE}(\xi) \equiv \exp(-\psi(\xi))$ where $\xi \equiv r/a$ is calculated from the Lane-Emden equation of isothermal gas with the boundary condition of
$\psi(\xi)=d\psi(\xi)/d\xi=0$ at $\xi=0$,
\begin{align}
\frac{d}{d \xi}\left(\xi^2 \frac{d \psi}{d \xi}\right)=\xi^2 \exp(-\psi).
\end{align}
}

A Bonnor-Ebert sphere is determined by specifying central density $\varrho_0$, the
ratio of the central density to density at $R_c$,  $\varrho_0/\rho_0(R_c)$.
In this study, we adopted the values of $\varrho_0=2.8\times 10^{-18} \gcm$,
$\varrho_0/\rho_0(R_c)=14$.
{ The enclosed mass within $R_c=5.3\times 10^3$ AU and $10^4$ AU are $0.53 \msun$ and $1 \msun$, respectively.}
We assume a rigid rotation with the angular velocity of  $\Omega_0=10^{-13} {\rm ~s^{-1}}$,
to be consistent with the observational results \citep[e.g.,][]{2011ApJ...740...45T}.

\begin{figure}
  \includegraphics[trim=0mm 0mm 0mm 0mm,width=70mm,angle=0]{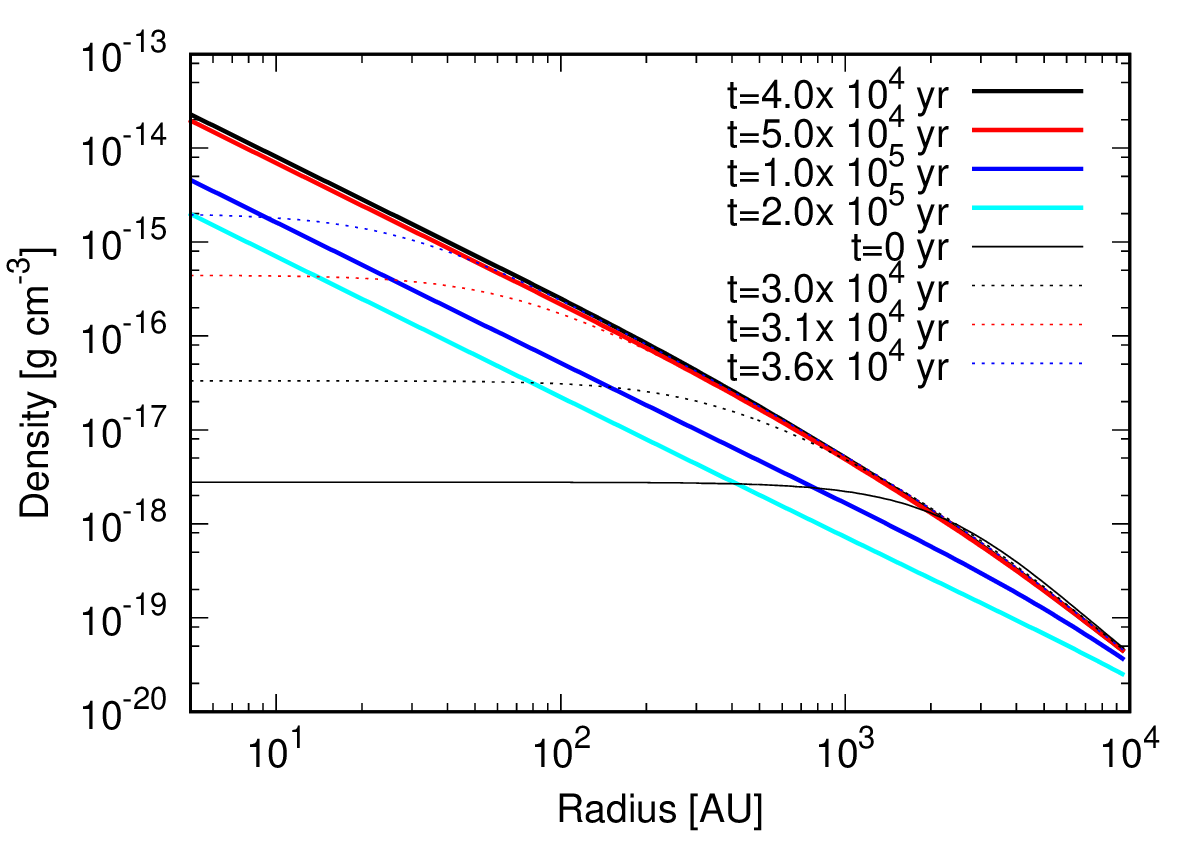}
  \includegraphics[trim=0mm 0mm 0mm 0mm,width=70mm,angle=0]{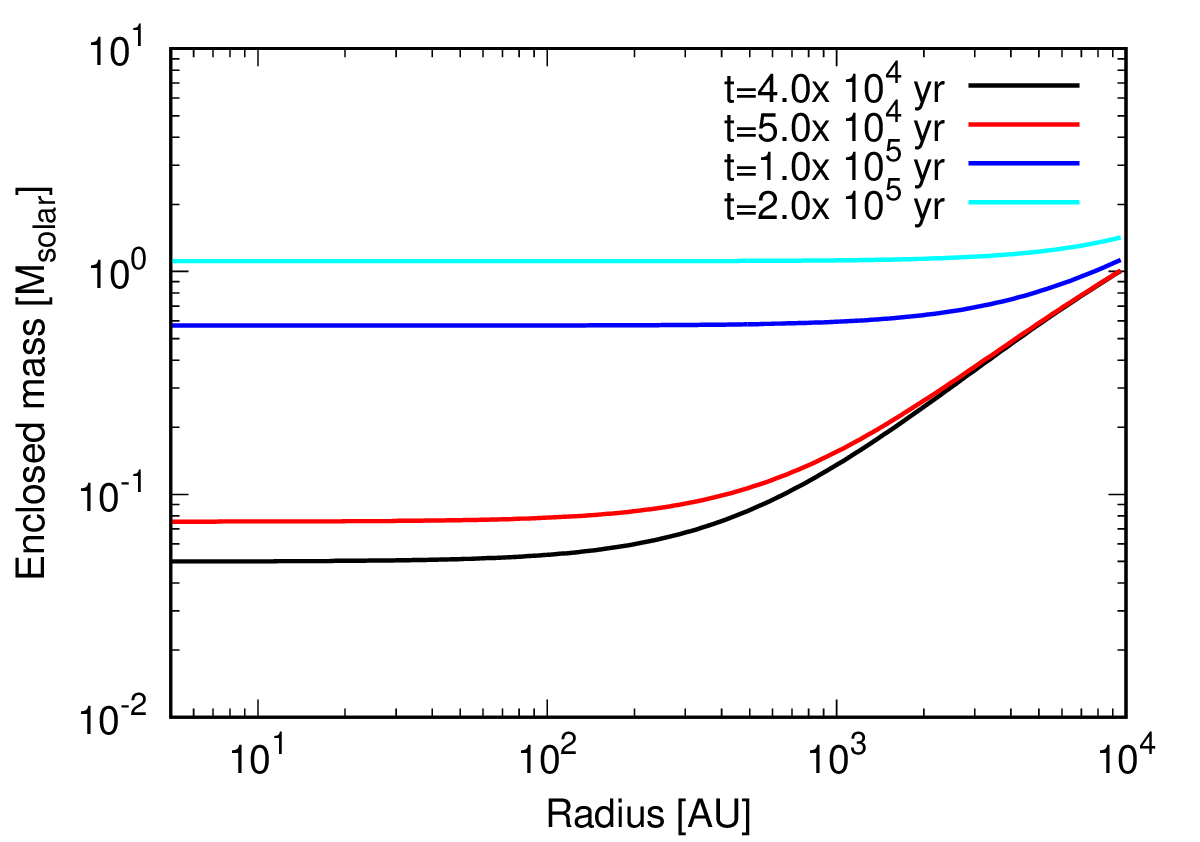}
  \includegraphics[trim=0mm 0mm 0mm 0mm,width=70mm,angle=0]{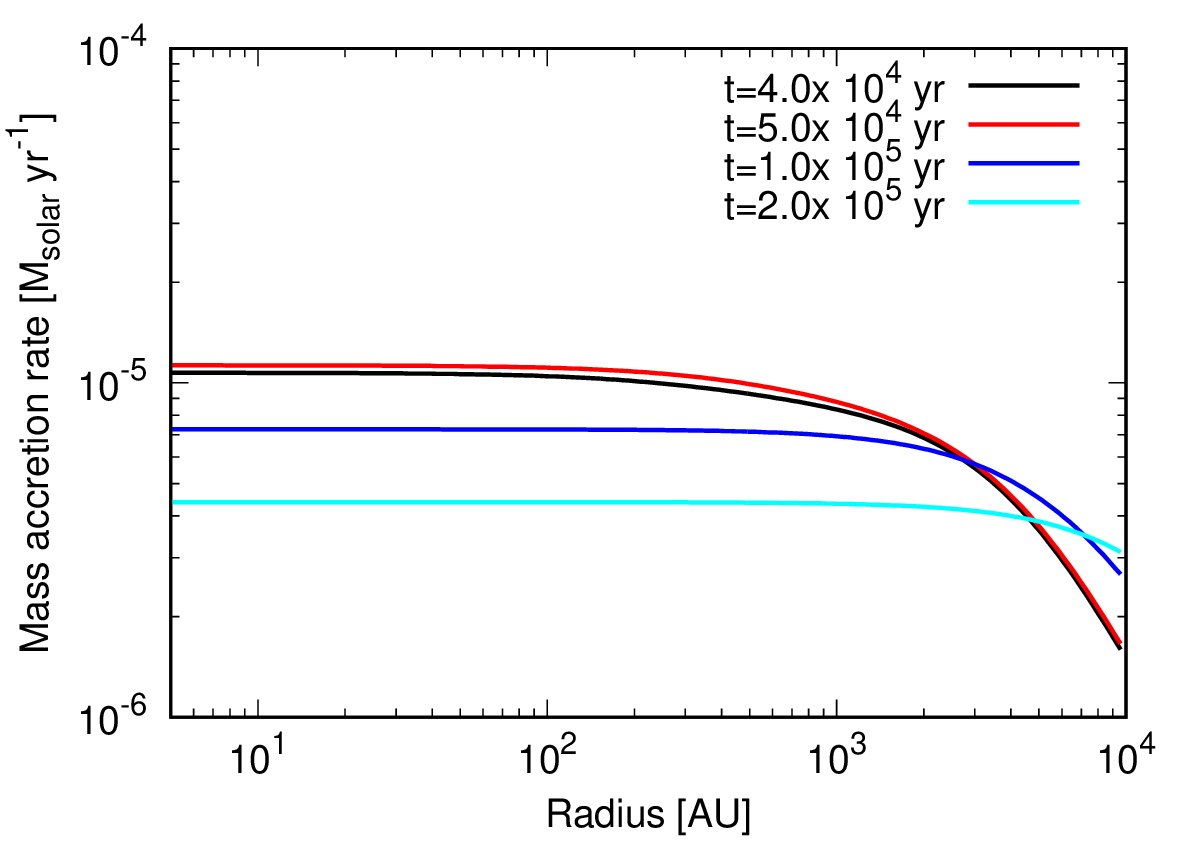}
  \includegraphics[trim=0mm 0mm 0mm 0mm,width=70mm,angle=0]{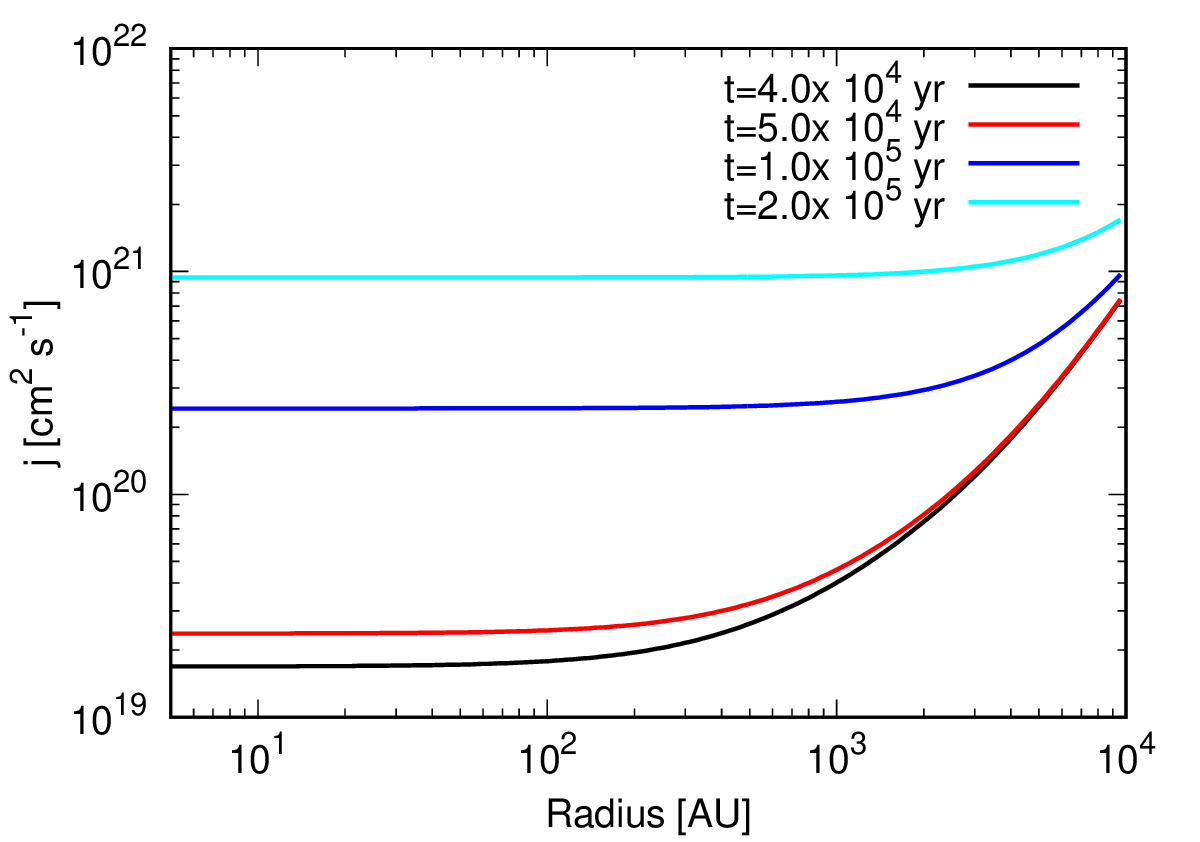}
  \caption{
    Radial profiles of density, enclosed mass, mass accretion rate, and specific angular momentum of the collapse model.
    Thick solid lines show profiles of late accretion phase and black, red, blue, and cyan lines show the profiles
    at $t=4\times 10^4, 5\times 10^4,  1\times 10^5,  2\times 10^5$ yr, respectively.
    Thin solid black line in density profile shows initial density.
    Dotted lines in density profile show the profiles of prestellar collapse phase and black,
    red, and blue lines show the profiles
    at $t=3.0\times 10^4, 3.1\times 10^4,  3.6\times 10^4$ yr, respectively.
}
\label{rad_prof_env}
\end{figure}

Figure \ref{rad_prof_env} shows the radial profile of density, enclosed mass, mass accretion rate, and specific angular momentum of our envelope model.
We can see that enclosed mass, mass accretion rate, and specific angular momentum are almost radially constant in $r \sim 100$ AU in late accretion phase (thick solid lines).
Thus, we use enclosed mass at $r=100$ AU as the total mass of central protostar and disk
\begin{align}
  M(r={\rm 100AU},t)= M_{\rm tot} 
\end{align}
the mass accretion rate at $r=100$ AU as the envelope-to-disk mass accretion rate,
\begin{align}
  \dot{M}_{\rm env}(r={\rm 100AU},t)= \dot{M}_{\rm env}
\end{align}
and the specific angular momentum  at $r=100$ AU as the specific angular momentum of accretion flow to the disk.
\begin{align}
  j(r={\rm 100AU},t)= j_{\rm env} 
\end{align}

\begin{figure}
  \includegraphics[trim=0mm 0mm 0mm 0mm,width=70mm,angle=0]{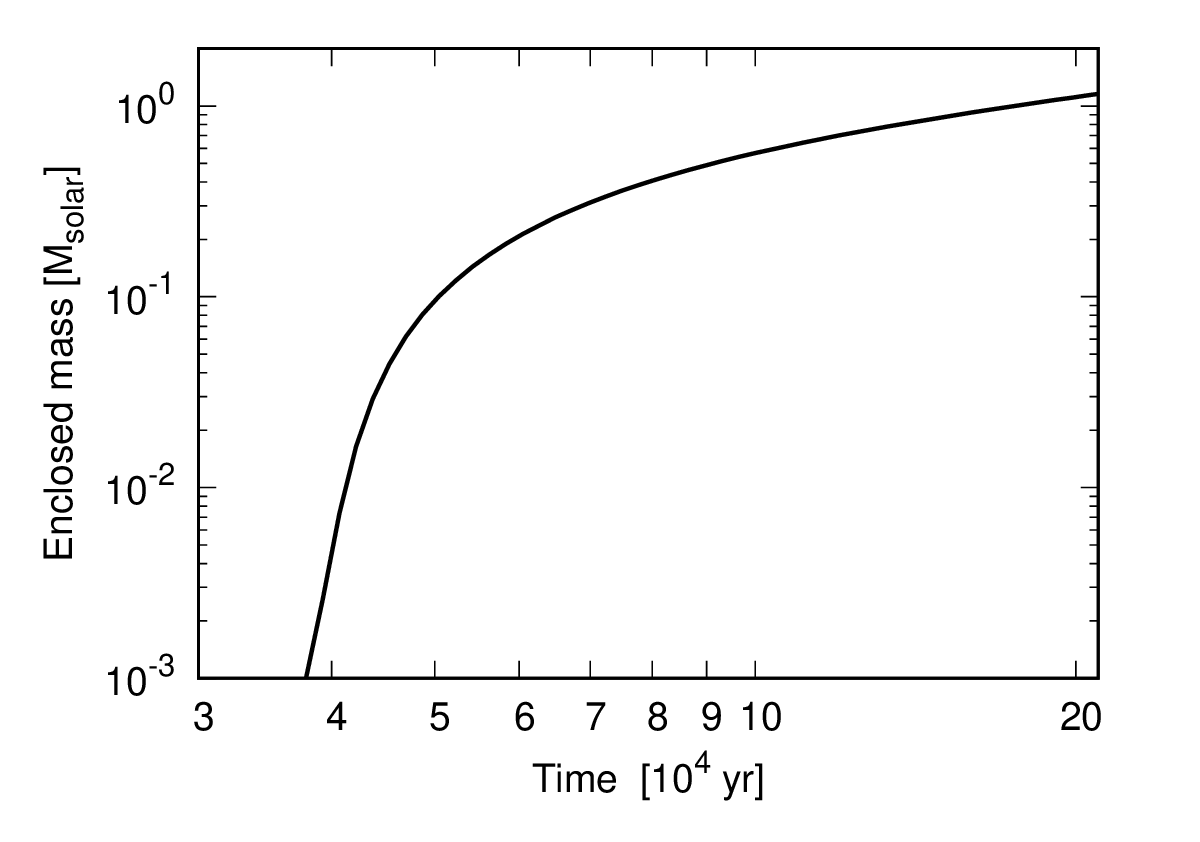}
  \includegraphics[trim=0mm 0mm 0mm 0mm,width=70mm,angle=0]{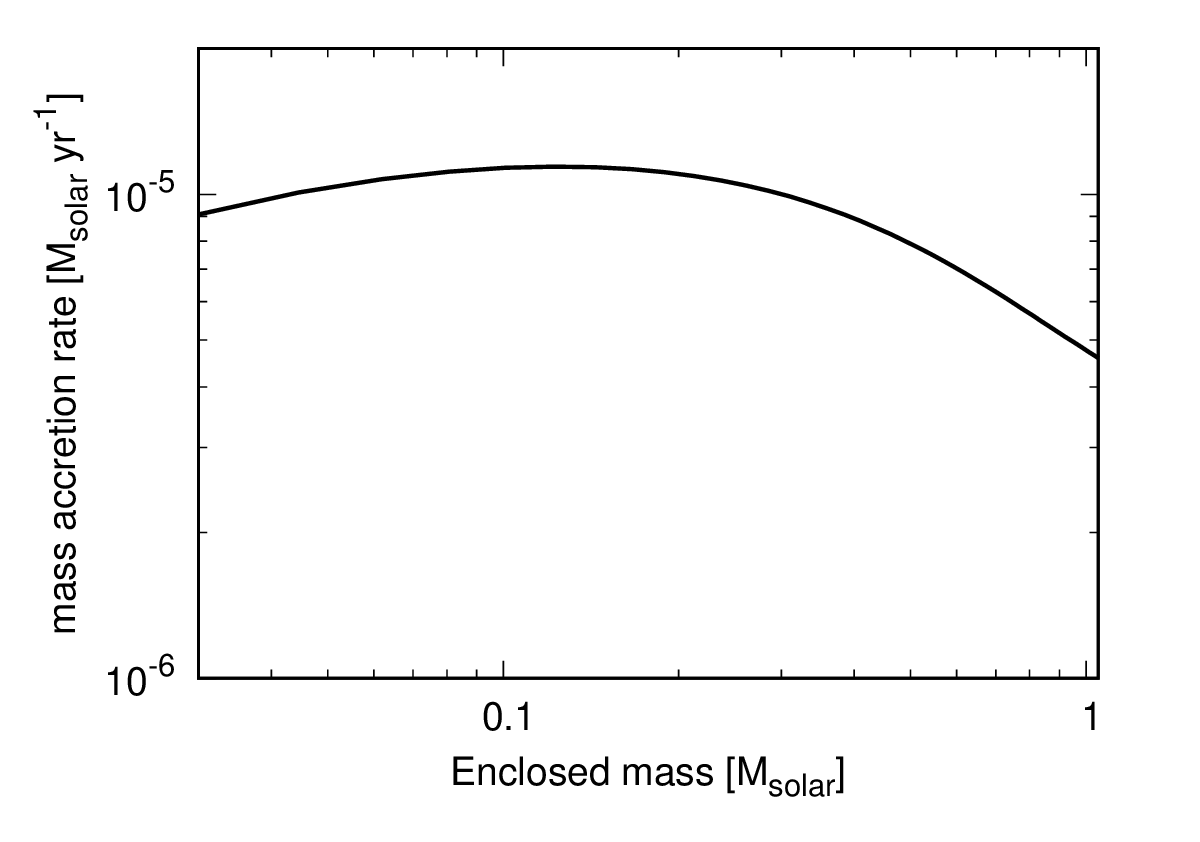}
  \includegraphics[trim=0mm 0mm 0mm 0mm,width=70mm,angle=0]{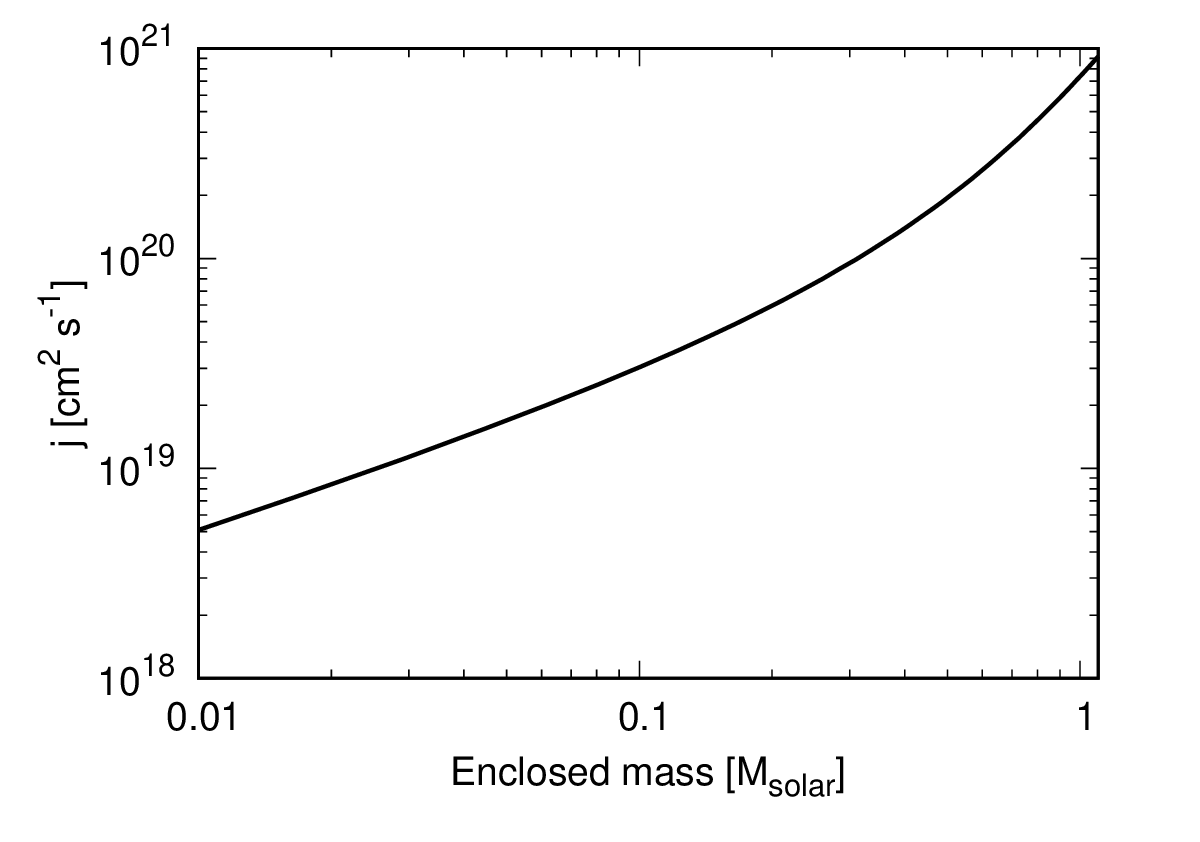}
  \caption{
    Time evolution of enclosed mass, and specific angular momentum of the collapse model at $r=100$ AU.
    The mass accretion rate and specific angular momentum are shown as a function of enclosed mass. 
}
\label{time_prof_env}
\end{figure}

Figure \ref{time_prof_env} shows the time evolution of enclosed mass, mass accretion rate, and specific angular momentum of our envelope model at $r=100$ AU.
In our collapse model, it takes $2\times 10^5$ yr for the total mass of the central star and disk $M_{\rm tot}$ to reach $\sim 1 \msun$.
During this period, the mass accretion rate reaches $\sim 10^{-5} \msunyr$ at $M_{\rm tot}=0.1 \msun$ and decreases to $\sim 4 \times 10^{-6} \msunyr$ at $M_{\rm tot}=1 \msun$.
On the other hand, the specific angular momentum of the accretion flows continues to increase monotonically and reaches $j_{\rm env} \sim 10^{21} \cmcms$ at $M_{\rm tot}=1 \msun$.

\subsection{Connection between the envelope model and the disk model}
Here, we explicitly specify the relationship between the quantities of the envelope model and those of the disk model.
To connect the accretion model described in the previous subsection with our disk model,
we assume that the total mass of central protostar and disk is given as
\begin{align}
  \label{totM}
  M_{\rm tot}=M_{\rm disk}+ M_{\rm star}, 
\end{align}
that the mass accretion rate from envelope to be disk mass accretion rate is given as,
\begin{align}
\dot{M}_{\rm disk}=\dot{M}_{\rm env},
\end{align}
and that the angular momentum accretion rate from envelope is equal to the angular momentum removal rate,
\begin{align}
  \label{j_balance1}
\dot{J}_{\rm out}=\dot{J}_{\rm env}=\dot{M}_{\rm env} j_{\rm env}.
\end{align}
Hereafter, we refer to this disk-envelope model as "co-evolution disk model".

On the other hand, we also consider a disk that evolves by internal angular momentum transport due to gravitational instability for comparison.
Hereafter, we refer to this model as "GI disk model".
In this GI disk model, we assume
\begin{align}
    \label{j_balance2}
\int^{r_{\rm disk}}_0 \Sigma_{\rm GI}~ r^2 \Omega ~2\pi r dr=\int^t_0 \dot{M}_{\rm env}(t) j_{\rm env}(t) dt,
\end{align}
instead of equation (\ref{j_balance1}) because total angular momentum in the disk is not removed by internal viscous accretion process.

To obtain disk mass and protostar mass,
we numerical solve equations (\ref{j_balance1}) (for co-evolution disk model)
or (\ref{j_balance2}) (for GI disk model) and equation (\ref{totM}) to obtain $M_{\rm disk}$ and $M_{\rm star}$.
Although the angular velocity of the disk can be different from the angular velocity of the Keplerian rotation
when the disk mass is not negligible compared to the central protostar mass, we simply assume
\begin{align}
\Omega=\sqrt{\frac{G M_{\rm star}}{r^3}},
\end{align}
even for the gravitationally unstable disk.

\subsection{Evolution of disk radius}
\begin{figure}
  \includegraphics[trim=0mm 0mm 0mm 0mm,width=70mm,angle=0]{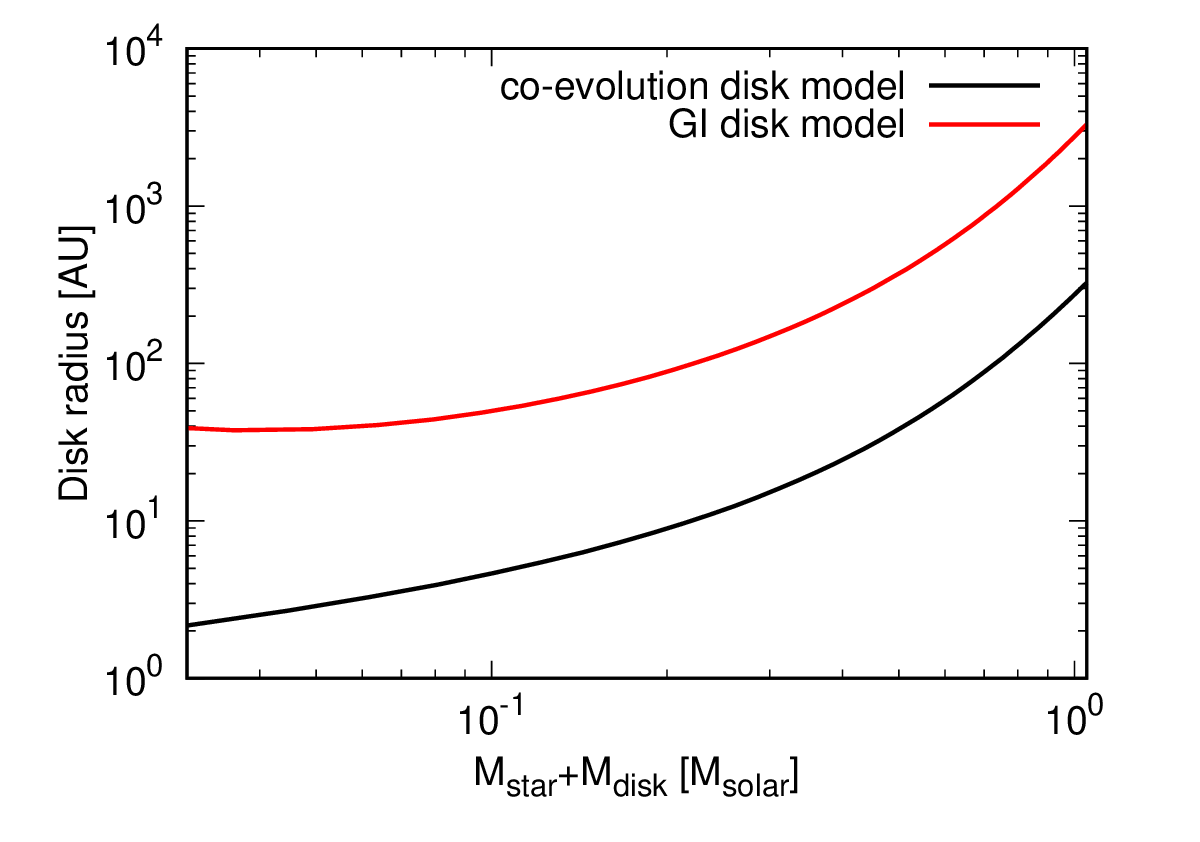}

  \caption{
    Time evolution of disk radius of the co-evolution model (black) and GI disk model (red).
    The temperature for GI disk model is assumed to be $T_0=150$ K.
}
\label{time_rad_env}
\end{figure}

Figure \ref{time_rad_env} shows the disk radius as a function of the total mass, which can be regarded as the time evolution of the disk radius.
The co-evolution model predicts the disk radius of several AU at $M_{\rm tot}\sim 10^{-1} \msun$ to several 100 AU
in $M_{\rm tot}\sim 1 \msun$. This relatively compact disk seems to be consistent with the recent observations
of Class 0/I YSOs \citep{2017ApJ...834..178Y,2019A&A...621A..76M}.

On the other hand, radius of GI disk model is almost 10 times larger than the co-evolution model
and several $ 10$ AU at $M_{\rm tot}\sim 10^{-1} \msun$ and reaches several 1000 AU at $M_{\rm tot} \sim1 \msun$.
This radius especially around $M_{\rm tot}\sim 1 \msun$ is very large and seems to be inconsistent with the observations.
Note that, the radius of GI disk is the {\it lower limit} among disks which evolves with the internal angular momentum transport,
because the surface density of GI disk is the upper limit and given total angular momentum, smaller surface density causes larger disk radius.
Thus, we conclude in general that disk evolution models in which disks evolve with internal angular momentum transport mechanisms
predict too large disk radius and at least a certain amount of angular momentum must be extracted from the disk during Class 0/I phase.

\subsection{Evolution of disk mass}
\begin{figure}
  \includegraphics[trim=0mm 0mm 0mm 0mm,width=70mm,angle=0]{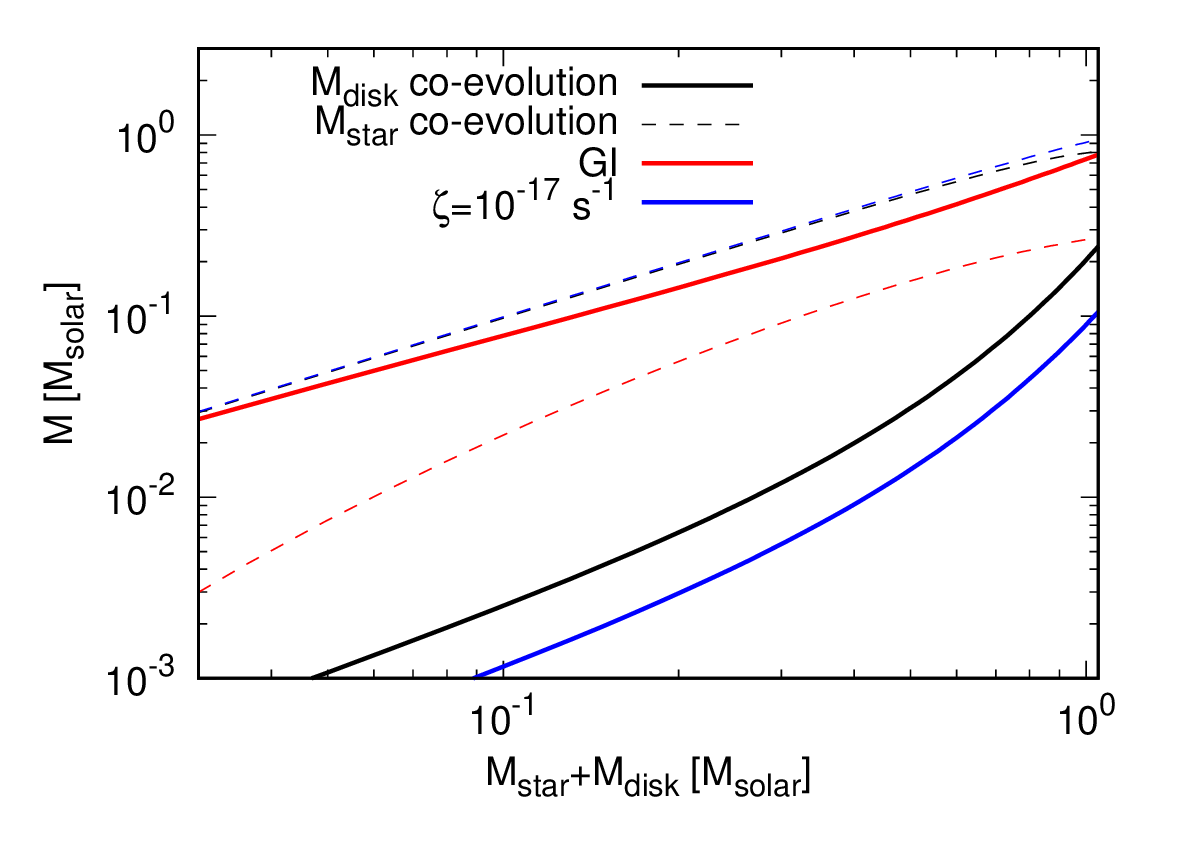}
  \caption{
    Time evolution of mass of disk (solid lines) and central protostar (dashed lines) of the co-evolution model (black, blue) and GI disk model (red).
    The temperature is assumed to be $T_0=150$ K. $\zeta$ for black and blue lines is $\zeta=10^{-18} {\rm ~s^{-1}}$, $\zeta=10^{-17} {\rm ~s^{-1}}$, respectively.
}
\label{time_mass_env}
\end{figure}

Figure \ref{time_mass_env} shows the evolution of the mass of disk and protostar.
The disk mass of co-evolution model predicts the disk mass of several $ 10^{-3} \msun$ at $M_{\rm tot}\sim 10^{-1} \msun$ 
and mass of $\sim 10^{-1} \msun$ at $M_{\rm tot}\sim 1 \msun$
(which depends on the ionization rate as $\zeta^{-1/3}$).
The disk mass is much smaller than the central protostar mass.
On the other hand, the disk mass of GI disk model is very large and it is larger than the central protostar mass, $M_{\rm disk}/M_{\rm star}\gtrsim 1$.
Although $M_{\rm disk}/M_{\rm star}\gtrsim 1$ is actually obtained in the hydrodynamics simulations starting from collapsing cloud core \citep[e.g.,][]{2011MNRAS.416..591T,2011PASJ...63..555M},
such a extremely massive disk seems to be inconsistent with the observations.

\subsection{Evolution of disk magnetic-flux}
\begin{figure}
  \includegraphics[trim=0mm 0mm 0mm 0mm,width=70mm,angle=0]{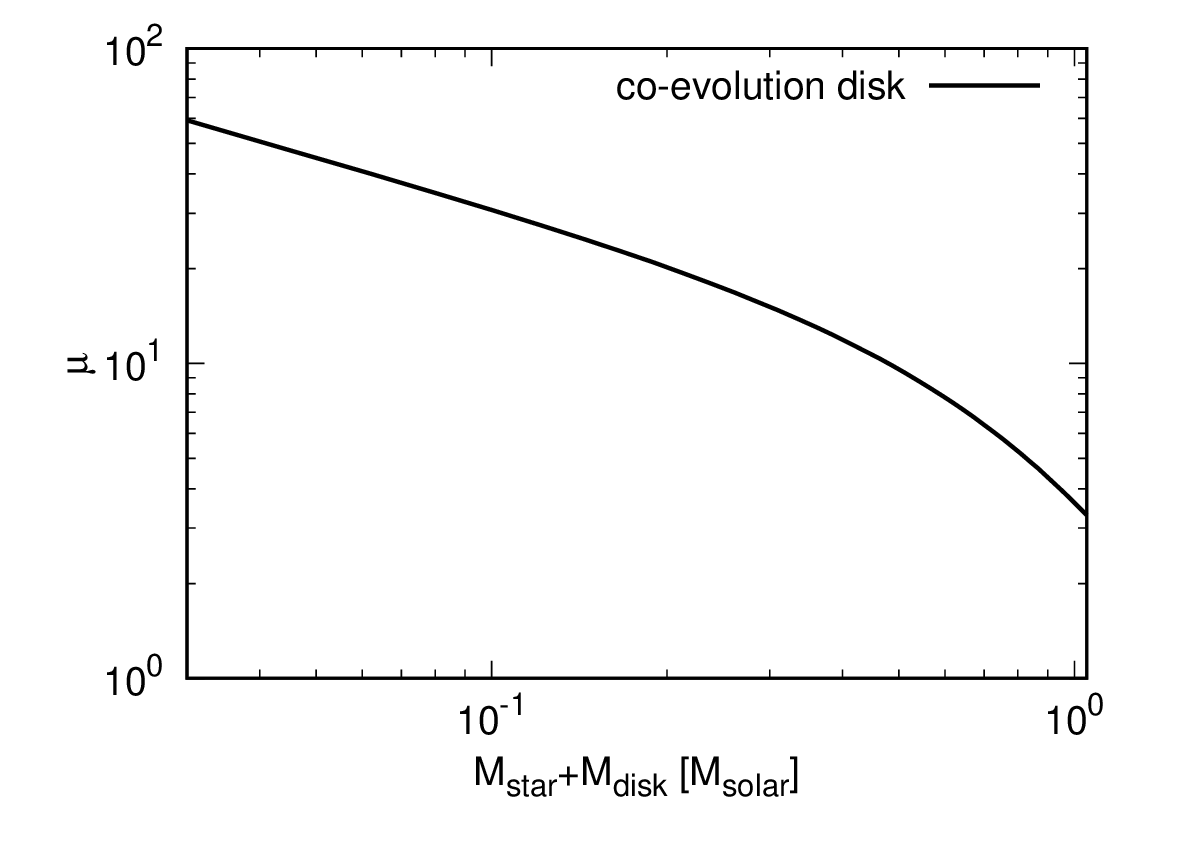}
  \caption{
    Time evolution of mass-to-flux ratio of the disk-protostar system normalized
    by $\left(\frac{M}{\Phi}\right)_{\rm crit}=\frac{0.53}{3 \pi}\left(\frac{5}{G}\right)^{1/2}$ of co-evolution model.
    The temperature and $\zeta$ are assumed to be $T_0=150$ K and  $\zeta=10^{-18} {\rm ~s^{-1}}$, respectively.
}
\label{M_mu_env}
\end{figure}

Figure \ref{M_mu_env} shows the time evolution of the mass-to-flux ratio $\mu$ of co-evolution model.
With this figure, we investigate whether the magnetic flux provided by envelope accretion
is sufficient to realize the vertical magnetic field of co-evolution disk model.
If the magnetic flux of the disk is larger than that of the initial molecular cloud core,
the vertical magnetic field of our disk cannot be realized and our model becomes inappropriate.

The figure shows that $\mu \sim 30$ at $M_{\rm tot}\sim 10^{-1} \msun$.
{ On the other hand, the mass-to-flux ratio of the cloud core is $\mu=O(1)$ \citep{2012ARA&A..50...29C}.
  Thus, it indicate that sufficient magnetic flux is supplied to the disk in the early evolutionary phase.}
$\mu$ monotonically decreases mainly due to the increase of the disk radius.
{
When $M_{\rm tot}\sim 1 \msun$ (and $r_{\rm disk} \sim 300$ AU with our envelope model),
$\mu \sim 3$ which is the same order of the mass-to-flux ratio of the cloud core.
Thus, if the disk grows to several $100$ AU, the disk can reserve the same level of magnetic flux of the cloud core.
Furthermore, if $\mu$ of the core is $\mu \gtrsim 3$,
the magnetic flux of the cloud core is not sufficient to realize vertical magnetic field of equation (\ref{Bz}).
Therefore, we must be careful about applying the disk model in this paper to model the outer region of several hundred AU sized disks.
}

\subsection{Specific angular momentum profile from disk to envelope}
So far, we have compared the co-evolution disk model with the GI disk model, and
we pointed out that the co-evolution disk model predicts the radius and mass of the disk more consistent with observations than the GI disk model.

{ Is there any way to more rigorously distinguish these two disk evolution models from observations?
(A more generalized question would be a way to distinguish the disk models in which the angular momentum removal balances the angular momentum supply
and disk models which evolve by the internal angular momentum transport mechanism.)}
We propose that these two (types of) models can be distinguished by examining
the radial profile of specific angular momentum (or rotational velocity) from the envelope to the disk.

For disks that evolve by  angular momentum transfer within the disk, such as viscous accretion disks,
the disk size is determined by the total angular momentum brought in during the accretion history,
since the total angular momentum of the disk is conserved (equation (\ref{j_balance2})).
In this case, the specific angular momentum of the accreting flow at a given instant
does not generally coincide with the specific angular momentum at the outer edge of the disk.
Therefore, there must be a jump in the specific angular momentum profile at the envelope-disk boundary.

On the other hand, in a steady accretion disk in which the angular momentum extraction from the disk is balanced by the angular momentum supply to the disk,
the specific angular momentum of the accretion gas and that
at the outer edge of the disk do coincide, as seen in equation (\ref{r_disk1}).
In this case, the angular momentum profiles are continuously connected at the envelope-disk boundary.
Therefore, we can determine whether the disk is evolving according to viscous evolution by MRI, GI, etc.
or by the magnetic braking from the existence of jumps in the specific angular momentum profiles.

\begin{figure}
  \includegraphics[trim=0mm 0mm 0mm 0mm,width=70mm,angle=0]{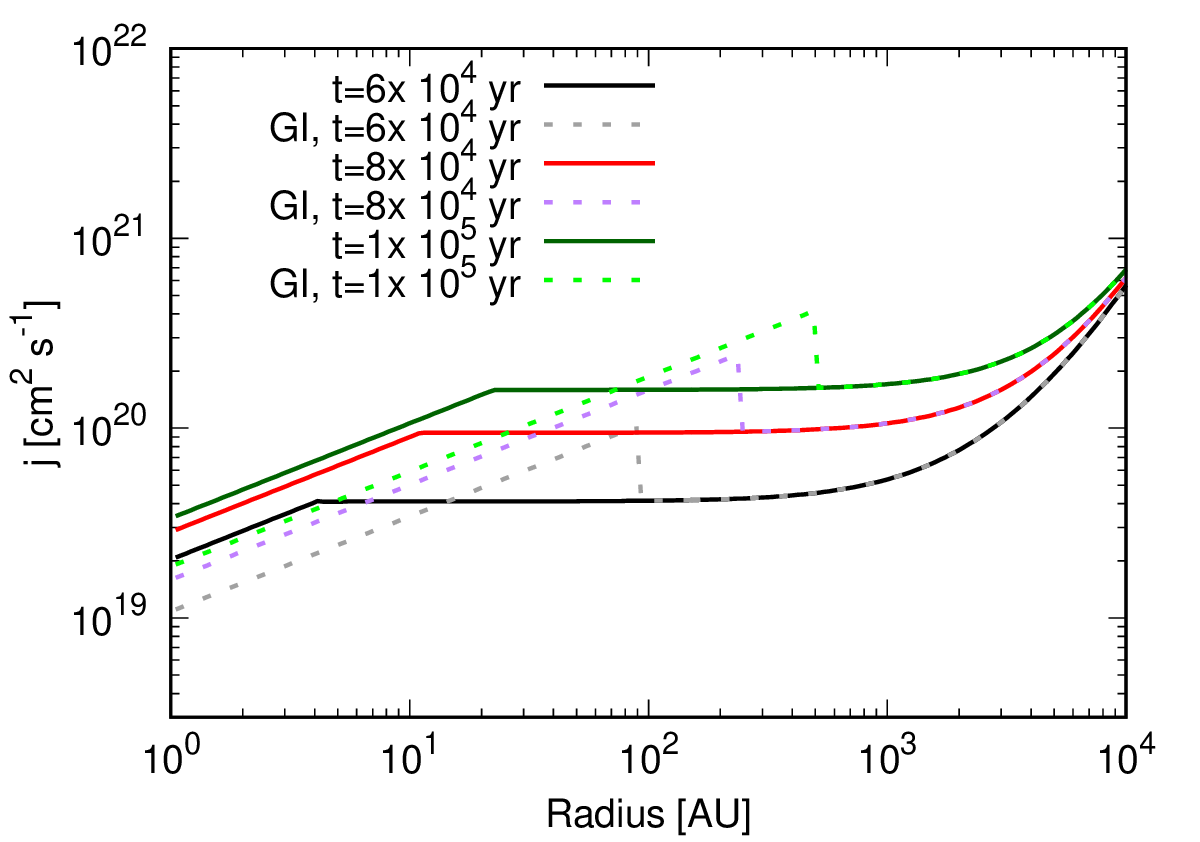}
  \includegraphics[trim=0mm 0mm 0mm 0mm,width=70mm,angle=0]{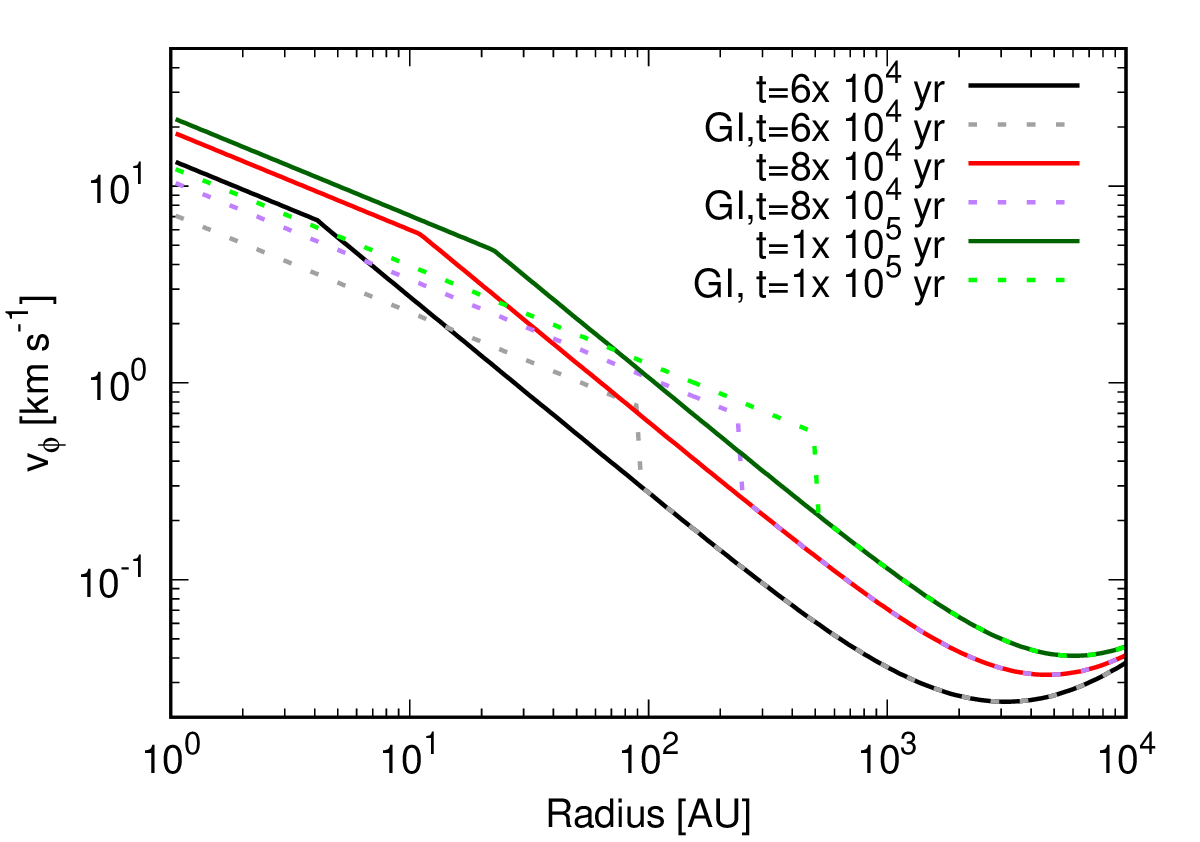}
  \caption{
    Radial profiles of specific angular momentum { and rotation velocity} of co-evolution model (solid) and GI disk model (dashed)
    at $t=6\times 10^4$ (black;co-evolution model, gray;GI model), $8\times 10^4$ (red;co-evolution model, purple;GI model),  $1\times 10^5$ yr
    (green;co-evolution model, yellow-green;GI model), respectively.
    The temperature and $\zeta$ are assumed to be $T_0=150$ K and  $\zeta=10^{-18} {\rm ~s^{-1}}$, respectively.
}
\label{r_j_prof_env}
\end{figure}

We show the radial profile of the specific angular momentum (and rotation velocity) of co-evolution disk model
(solid lines) and GI disk model (dashed lines)  in figure \ref{r_j_prof_env}.
In the co-evolution disk model, the specific angular momentum continuously transit from $j \propto r^0$ to $j \propto r^{1/2}$ 
{ (or equivalently, the rotation velocity continuously transit from $v_{\rm \phi} \propto r^{-1}$ to $v_{\rm \phi} \propto r^{-1/2}$)} without jump.
On the other hand, in the GI disk, the specific angular momentum increases about a factor of two at the boundary between the disk and the envelope throughout
the entire accretion phase.
{ The difference of the disk rotation profiles between two models at the same epoch are due to the difference of the protostar masses (we assume $\Omega=\sqrt{GM_{\rm star}/r^3}$).}
Because the specific angular momentum profile in the disk depends weakly on radius as  $j \propto r^{1/2}$,
the jump if it exists would be observable with the observations in which the disk is spatially resolved.

{ Note that the magnitude of the jump of GI disk in the figure is a conservative estimate.
  In the GI disk, the mass of the central star is much smaller than the mass of the disk, but we only consider the mass of the central star to calculate $\Omega$.
  In reality, at the outer edge of the GI disk, the radial self-gravity of the
  disk may cause the faster angular velocity
  (or shallower radial profile of $\Omega$ than Keplerian rotation profile)
  \citep[see, e.g.,][]{2015MNRAS.446.1175T}.
  In this case, the magnitude of the jump is expected to be larger.
  It is noteworthy that there is a factor of two increase even with such a conservative estimate.
}

\section{Summary and Discussion}
\label{discussion}
\subsection{Summary}
In this paper, we analytically investigate the disk structure predicted from "co-evolution disk model".
In our disk model, we assume that the dust grains have grown sufficiently and ambipolar resistivity is
determined by the balance between ionization and gas-phase recombination.
We also assume that the disk evolves with magnetic braking rather than internal viscous accretion.
We utilize the equations (\ref{eq1}) to (\ref{eq6}) which can be regarded as the modified steady accretion disk model,
and the equation (\ref{model_etaA}) which describes the sufficient dust growth in the disk, to obtain the disk structure.
We have confirmed in our previous paper TMI23 that these equations describe the disk structures obtained by 3D non-ideal MHD
simulations with dust growth very well (within a factor of three).

Our disk structure is shown by equation (\ref{rho}) to (\ref{etaA}),
and is described as a function of central stellar mass, mass accretion rate,
gas phase recombination rate, gas temperature, and cosmic ray ionization rate, and does not include free-parameter such as viscous parameter $\alpha$.
With this disk model, we discussed whether GI and MRI can develop or not during Class 0/I phase.
Then, by combining an analytical model of envelope accretion  with the disk model, we made predictions for the disk mass and radius during Class 0/I phase.

Our main findings are summarized as follow.
\begin{itemize}
\item Disk radius is estimated to be several AU to $\sim 100$ AU during the accretion phase which is about 10 times smaller than the disk evolving with angular momentum conservation.
\item Disk mass is estimated to be several $10^{-3} \msun$ to $\sim 10^{-1} \msun$ during the accretion phase which is consistent with the estimate from the observations.  
\item With typical disk ionization rates ($\zeta \gtrsim 10^{-18}  {\rm ~s^{-1}}$) and moderate mass accretion rate ($\dot{M}_{\rm disk}>10^{-8} \msunyr$), magneto-rotational instability is prohibited in the disk.
\item Plasma $\beta$ at midplane depends on the mass accretion rate (i.e., disk evolutionary stage) as $\beta \propto \dot{M}^{-2/3}_{\rm disk}$, and $\beta \sim 30$ at $\dot{M}_{\rm disk} \sim 10^{-6} \msunyr$ and $\beta \sim 700$ at $\dot{M}_{\rm disk} \sim 10^{-8} \msunyr$ at $r=10$ AU. 
\item Specific angular momentum (or rotation velocity) profile at the envelope to disk boundary is continuous if the disk angular momentum evolves only with magnetic braking.
\end{itemize}

\subsection{Discussion}
\subsubsection{The relation of the disk size and the specific angular momentum of accretion flow}
An important prediction obtained from our analysis of the disk model is that the disk radius coincides
with the centrifugal radius of the accreting flow when the disk evolves with magnetic braking rather than viscous accretion.
This follows just from the assumption of steadiness of the disk (i.e., constant $\dot{M}_{\rm disk}$)
and the assumption that angular momentum extraction from the disk is balanced with angular momentum supply to the disk,
and does not depend on the detail of the assumptions on $\eta_A$.

This prediction can be verified by detailed observations
of the specific angular momentum (or rotation velocity) profile around the disk-envelope boundary.
If the specific angular momentum is continuous at the disk-envelope boundary, it suggests that disk angular momentum
evolution is determined by the angular momentum extraction due to the magnetic braking.
On the other hand, if the specific angular momentum is discontinuous (or has the jump) at the disk-envelope boundary,
it suggests that disk angular momentum evolution is determined by the internal (viscous) angular momentum transport (see figure \ref{r_j_prof_env}).

There have been several observations that investigate the radial profile of specific angular momentum or rotation velocity of Class 0/I YSOs.
Recent ALMA velocity profile observations seem to show that 
rotational velocity profile from the envelope to disk of the Class 0/I phase seems to be continuous and no jump
at the disk-envelope boundary \citep{2014ApJ...796..131O,2015ApJ...812...27A,2017ApJ...834..178Y}.
If future high-resolution observations confirm the continuous connection of the specific angular momentum
of the envelope and the outer edge of the disk, this indicates that the evolution of the angular momentum and radius of the disk is determined
by the extraction of angular momentum from the disk (i.e., magnetic braking) rather than by internal viscous evolution of the disk.

\subsubsection{Caveat of co-evolution disk model and Future Prospects}
In the present study, ambipolar diffusion is assumed to be the most efficient non-ideal MHD effect.
When dust growth proceeds, it is shown that Ohmic dissipation and Hall effects are weaker than ambipolar diffusion
for cases where the magnetic field is relatively strong \citep{2022ApJ...934...88T}.
However, when the magnetic field becomes weak, these effects may possibly play the role.
In future studies, we will examine the disk structure where Ohm dissipation or the Hall effect play the dominant role.
Note, however that, when Ohm dissipation is dominant, there may not be a unique steady structure because $\eta_O$ is independent of the magnetic field.

In this study, we assumed that the dust grans were well grown over the entire disk.
However, at the edge of the disk, it is expected that there is a supply of $\mum$ sized dust grains from the envelope.
In that case, more efficient magnetic diffusion would be expected at the edge of the disk, and our approximation for $\eta_A$ may not be justified.
It is important to study the physical state of the outer edge of the disk in detail in the future studies which consider the dust growth and envelope-to-disk accretion \citep[e.g.,][]{2017ApJ...838..151T}.
Note, however, that the above discussion of the specific angular momentum jump at the outer edge of the disk
is not affected by the dust size as long as it evolves with magnetic braking, since it does not depend on the details of $\eta_A$.

{
  We also note that the detailed modeling of dust coagulation and fragmentation
  and its impact on $\eta_A$
  is important because the dust fragmentation can cause a variety of dust size distributions \citep{2011A&A...525A..11B}.
  As shown in \citet{2023MNRAS.518.3326L},
  dust fragmentation can changes $\eta_A$ profile especially in the high density region.
  Once a model of $\eta_A$ that takes into account the effects
  of dust coagulation/fragmentation is obtained,
  it would be possible to predict the disk structure
  according to the dust growth model and resulting $\eta_A$ using the equations in this paper.
}

Although there are issues to be solved in the future as described above,
we believe that our co-evolution disk model provides a new perspective on the structure and evolution of protoplanetary disks.

\section*{Acknowledgments}
I thank Dr. Shinsuke Takasao, Dr. Satoshi Okuzumi, and Mr. Ryoya Yamamoto for the fruitful discussions.
I also thank Prof. Shu-ichiro Inutsuka and Prof. Masahiro Machida for their continuous encouragement and the comments.
We also thank the referee, Dr. Pierre Marchand for his helpful and insightful comments. 
This work is supported by JST FOREST.


\bibliography{article}

\end{document}